\documentclass{emulateapj}

\usepackage{multirow}
\usepackage{amsmath}

\newcommand{\beq}	{\begin{equation}}
\newcommand{\eeq}	{\end{equation}}
\newcommand{\beqa}{\begin{eqnarray}}
\newcommand{\eeqa}{\end{eqnarray}}
\newcommand{\beqs}	{\begin{displaymath}}
\newcommand{\eeqs}	{\end{displaymath}}
\newcommand{\beqas}	{\begin{eqnarray*}}
\newcommand{\eeqas}	{\end{eqnarray*}}

\def\bit{\begin{itemize}}
\def\eit{\end{itemize}}
\newcommand{\e}	{$^{-1}$}
\newcommand{\ee}	{$^{-2}$}
\newcommand{\eee}	{$^{-3}$}
\def\simlt{\lower.5ex\hbox{$\; \buildrel < \over \sim \;$}}
\def\simgt{\lower.5ex\hbox{$\; \buildrel > \over \sim \;$}}
\def\la{\simlt}
\def\ga{\simgt}

\font\tenbi=cmmib10 
\newfam\bifam  \textfont\bifam=\tenbi

\font\tenbr=cmbx10
\newfam\brfam  \textfont\brfam=\tenbr

\font\squinttenbi=cmbx10 at 9pt
\scriptfont\brfam=\squinttenbi

\def\vecnabla{
              \setbox1=\hbox{$\bigtriangledown$}
                           \raise.45ex\hbox{$\bigtriangledown$\hskip-.97\wd1
                           $\bigtriangledown$\hskip-.97\wd1
                           $\bigtriangledown$\hskip-.97\wd1}
                           \raise.47ex\hbox{$\bigtriangledown$}}

\def\symbol#1{\ifmmode#1\else$#1$\fi}

\newcommand{\pbyp}[1]	{{{\partial\hfil}\over{\partial#1}}}
\newcommand{\ppbyp}[2]	{{{\partial#1}\over{\partial#2}}}

\def\citeapos#1{\citeauthor{#1}'s (\citeyear{#1})}

\newcommand{\mda} 	{\dot{m}_\mathrm{acc}}

\newcommand{\mdw}	{\ppbyp{m_w(\varpi)}{t}}
\newcommand{\mdotw}  {\dot m_w}

\newcommand{\rhoch}	{\rho_{\rm ch}}

\newcommand{\rmax}	{R_{\rm max}}

\newcommand{\scl}	{\Sigma_{\rm cl}}

\newcommand{\twe}	{\theta_{w,\,\rm esc}}

\newcommand{\vkc}	{v_{Kc}}
\newcommand{\vp}	{\varpi}
\newcommand{\vpc}	{\varpi_c}
\newcommand{\vpd}	{\varpi_d}
\newcommand{\vpco}	{\varpi_{c0}}
\newcommand{\vpm}	{\varpi_{\rm max}}
\newcommand{\vpmo}	{\varpi_{\rm max,\,0}}
\newcommand{\vpo}	{\varpi_0}
\newcommand{\vzo}	{v_{z0}}

\newcommand{\vzco}	{v_{zc0}}

\newcommand{\xmo}	{x_{\rm max,\,0}}
\newcommand{\zdm}	{z_{d,\,\rm max}}

\newcommand{\thesc}	{\theta_{w,\,\rm esc}}

\shorttitle{Radiation Transfer of Models of Massive Star Formation II}
\shortauthors{Zhang, Tan \& McKee}

\begin{document}

\title{Radiation Transfer of Models of Massive Star Formation. II. Effects of the Outflow}

\author{Yichen Zhang}
\affil{Department of Astronomy, University of Florida, Gainesville, FL 32611, USA;\\yczhang@astro.ufl.edu}
\author{Jonathan C. Tan}
\affil{Departments of Astronomy \& Physics, University of Florida, Gainesville, FL 32611, USA;\\ jt@astro.ufl.edu}
\author{Christopher F. McKee}
\affil{Departments of  Physics \& of Astronomy, University of California, Berkeley, CA 94720, USA;\\ cmckee@berkeley.edu}

\begin{abstract}
We present radiation transfer simulations of a massive ($8 M_\odot$)
protostar forming from a massive ($M_{\rm c}=60 M_\odot$) protostellar
core, extending the model developed by \citet[Paper I]{ZT11}.  The two
principal improvements are (1) developing a model for the density and
velocity structure of a disk wind that fills the bipolar outflow
cavities, based in part on the disk-wind model of \citet[]{BP82}; and
(2) solving for the radially varying accretion rate in the disk due to
a supply of mass and angular momentum from the infall envelope
and their loss to the disk wind. 
One consequence of the launching of the disk wind is a reduction in the amount of accretion power that is radiated by the disk.
We also include a
non-Keplerian potential appropriate for a growing, massive disk. For
the transition from dusty to dust-free conditions where gas opacities
dominate, we now implement a gradual change as a more realistic
approximation of dust destruction.  We study how the above effects,
especially the outflow, influence the spectral energy distributions
(SEDs) and the synthetic images of the protostar. Dust in the outflow
cavity significantly affects the SEDs at most viewing angles. It
further attenuates the short-wavelength flux from the protostar,
controlling how the accretion disk may be viewed, and contributes a
significant part of the near- and mid-IR fluxes.  These fluxes warm
the disk, boosting the mid- and far-IR emission.  We find that for
near face-on views, i.e. looking down the outflow cavity (although not
too close to the axis), the SED from the near-IR to about 60 $\mu$m is
very flat, which may be used to identify such systems. We show that
the near-facing outflow cavity and its walls are still the most
significant features in images up to 70~$\mu$m, dominating the mid-IR
emission and determining its morphology.  The thermal emission from
the dusty outflow itself dominates the flux at $\sim20 \mu$m.  The
detailed distribution of the dust in the outflow affects the
morphology, so the model can be constrained by considering detailed
intensity profiles along and perpendicular to the outflow axis.  For
example, even though the outflow cavity is wide, at 10 to 20 $\mu$m,
the dust in the disk wind can make the outflow appear narrower than in
the near-IR bands. 

\end{abstract}


\section{Introduction}
\label{sec:intro}

There is still no consensus on how massive stars form. Possible
mechanisms include
collapse from massive cores supported by
turbulence and magnetic fields, as in the Turbulent
Core model (\citealt[]{MT03}, hereafter MT03), competitive accretion
at the centers of forming star clusters (\citealt[]{Bonnell01}), and
stellar collisions (\citealt[]{Bonnell98}). Theoretically, if a
massive star forms similarly to its low-mass counterpart, i.e. from a
core with a typical ratio of rotational to gravitational energy of a
few percent (\citealt[]{Goodman93}), a massive accretion disk with a
size of several hundreds to $\sim$ 1000 AU and bipolar outflows are
expected to appear around the massive protostar
(\citealt[]{Tan05}). Such disks and outflows have been seen in
hydrodynamic simulations with varying implementations of magnetic
field, turbulence and radiation (e.g., \citealt[]{KKM07},
\citealt[]{Kuiper10}, \citealt[]{Seifried11,Seifried12},
\citealt[]{Peters11}, \citealt[]{Vaidya11}). However, resolving this
issue observationally is difficult as massive stars tend to form in
distant, crowded, and highly obscured environments. Hot cores with
collimated molecular outflows and disk-like structures have been found
around massive young stars, but so far these observed disk-like
structures (or toroids) are much larger and more massive than expected
for the rotationally-supported disk of an individual protostar, and
there is little direct evidence for Keplerian or near-Keplerian disks
(e.g., \citealt[]{Cesaroni07}). The outflows have also been observed
in near- and mid-IR continuum emission, and it has been argued that
the outflow cavity may dominate the mid-IR morphology
(\citealt[]{Debuizer06}).

In order to better understand the properties of these observed massive
young stellar objects (MYSOs), a number of radiative transfer 
models have been developed to compare with observations. One of the
most popular is the model grid by \citet[]{Robitaille06}, but it is
focussed on low-mass star formation. Massive star-forming cores
usually reside in clumps within giant molecular clouds with large mass
surface densities, $\Sigma_\mathrm{cl} \sim 1\:{\rm g\:cm^{-2}}$
(MT03), so they are themselves relatively dense and collapse with high
accretion rates. The expected properties of the fiducial cores of MT03
lie outside the parameter space of the \citet[]{Robitaille06}
grid. Also the components in this radiative transfer
model are relatively simple --- e.g.,
the density in the outflow is usually assumed to be either a constant
value or a power-law. \citet[]{Offner12} compared this model grid with
results of numerical simulations of low-mass star formation, and found
possible discrepancies between the parameters inferred from fitting
with the radiative-transfer model grid and those of the protostars in the simulation,
especially for deeply embedded objects. Other radiative-transfer model grids, such as
those of \citet[]{Molinari08}, are for MYSOs, but the components are
also relatively simple. So it is worth including a more realistic
density distribution in the outflow cavity and constructing a fully
self-consistent model for all the components of the star-forming core.

This is the second of a series of papers on modeling the spectral
energy distributions (SEDs) and images of massive star-forming
cores. In the previous paper (\citealt[]{ZT11}, hereafter Paper I), we
studied a core with an initial mass of 60 $M_\odot$ that contained an
8 $M_\odot$ protostar forming at the center, and showed the effects of
an active accretion disk and an empty bipolar outflow cavity on the
SEDs and the images. The stellar mass chosen is high enough that the
star has a high luminosity ($L=2.8\times 10^3 L_\odot$) but low enough
that the ionizing luminosity is small; we ignore the latter here. We
included gas opacities inside the dust destruction front in the disk
since hot gas becomes optically thick. However, the outflow cavities
in Paper I were left empty (an optically thin limit), which is not
realistic since dust is expected to survive in these outflows;
furthermore, under certain conditions of high mass outflow rates,
the gas in the outflow may become optically thick near
the star.  In this paper, we shall study the effects on the SED of the
gas and dust in the wind filling the outflow cavities.  We shall also
consider how the outflow removes angular momentum and accretion energy
from the disk.  In the next section, we discuss our model setup. We
present the SED and image results of our models in Section
\ref{sec:result}, and summarize our main conclusions in Section
\ref{sec:summary}.  In future papers, based on the fiducial model
developed here, we shall present the SEDs and images of an
evolutionary sequence of MYSOs, and also examine their dependence on
the initial conditions of the core.

\section{Models}
\label{sec:models}

We now briefly describe the model setup in our previous work, which we
continue to use here. For a more detailed description of the model we
refer readers to Paper I. Following MT03, a star-forming
core is defined as a region of a molecular cloud that forms a single
star or a close binary via gravitational collapse. The core is assumed
to be spherical, self-gravitating and in near virial equilibrium. In
the fiducial case presented here, we study a core with an initial mass
of $M_\mathrm{c}=60 M_\odot$. The core is surrounded by a larger,
self-gravitating clump with a mean surface density
$\Sigma_\mathrm{cl}=1$ g cm$^{-2}$, which corresponds to pressure at
the surface of the core of
$\overline{P}_\mathrm{cl}=0.88G\Sigma_\mathrm{cl}^2
= 5.9\times 10^{-8} (\Sigma_\mathrm{cl}/{\rm g\:cm^{-2}})^2$~dyn~cm\ee. 
Such a core has a radius $R_\mathrm{c}=5.7\times
10^{-2}(\Sigma_\mathrm{cl}/{\rm g\:cm^{-2}})^{-1/2}$ pc ($1.2\times
10^4$ AU in the fiducial case).  We also consider two variants of this
model in which the surface density is smaller or larger than the
fiducial value by a factor of 0.5 dex. We assume that the initial
density distribution in the core is a power-law in spherical radius,
$\rho\propto r^{-k_\rho}$, and adopt $k_\rho=1.5$.
Such structures have recently been confirmed
to be present in Infrared Dark Clouds (IRDCs) by \citet[]{BT12}. Here
we study such a fiducial core at the particular moment when the
growing protostar reaches 8 $M_\odot$. Such a protostar has a radius
$r_*= 12.05\,R_\odot$ and a nuclear luminosity $L_*=2.81 \times
10^3\,L_\odot$ (MT03).

The collapse of the core is described with an inside-out expansion
wave solution (\citealt[]{Shu77}, \citealt[]{MP97}). The
expansion wave front moves outward, causing the material inside of it
to collapse; the outer part of the core remains static. We use
the solution by \citet[hereafter the Ulrich solution]{Ulrich76} to
describe the effects of rotation on the velocity field and the density
profile inside the sonic point, where the infall becomes supersonic. In
this solution, angular momentum is conserved until the material
falls onto the midplane and forms an accretion disk. Assuming the
rotational energy of the core is 2\% of its gravitational energy
($\beta=E_\mathrm{rot}/|E_\mathrm{grav}|=0.02$), the radius of the
disk $\varpi_d$ is $4.5\times 10^2$ AU in the fiducial case 
(Eq. 13 in Paper I).

Given a sufficiently large effective viscosity, the disk transfers
angular momentum outwards and enables the high accretion rates
required to form a massive star. The disk around a massive protostar
can have a high mass. 
In all our models with disks, 
we assume the mass ratio between the disk and the protostar is a
constant ($f_d=m_d/m_*=1/3$), 
motivated by an expected rise in effective viscosity due to disk
self-gravity at about this value of $f_d$
(e.g. \citealt[]{Adams89}, \citealt[]{Shu90}, \citealt[]{Gammie01}).
As in the previous paper, the disk structure is described with an
``$\alpha$-disk'' solution (\citealt[]{SS73}), but we have generalized
this to include the effects of the outflow and the accretion infall to
the disk (Section \ref{sec:disk}).

The outflow sweeps up part of the envelope and thus controls the
efficiency with which the star forms out of the core. We use the same
basic setup as in Paper I, but we have reduced the total mass loading
rate of the wind relative to the stellar accretion rate to
$f_w=\dot{m}_w/\dot{m}_*=0.1$ (rather than 0.6 in Paper I), which is a
typical value for disk wind models (\citealt[]{KP00}, also see Section
\ref{sec:outflow}). For better comparison to Paper I, we keep the
opening angle of the outflow cavity to be 51$^\circ$.  These
assumptions then lead to an increased star formation efficiency of
$\epsilon_{*f}=0.64$ (rather than 0.5 of Paper I). The major
difference from Paper I, where we treated the outflow cavity as being
completely optically thin, is that we have developed a theoretical
wind solution to evaluate a more realistic distribution of gas and
dust in the outflow cavity, which will be discussed in Section
\ref{sec:outflow}.

\subsection{Disk}
\label{sec:disk}

In Paper I, we considered a standard Shakura-Sunyaev disk with a
constant accretion rate
and a zero torque inner boundary condition. However, we will now
consider several additional effects that may be important in
protostellar accretion disks:
1) the accretion rate varies with radius because material in the
outflow (or the inflow) leaves (or joins) the disk; 2) similarly,
outflows and inflows alter the angular momentum distribution in the
disk; 3) the disk itself is massive and contributes to the
gravitational potential; and 4) the potential changes with time due to
the growth of the disk and the central object.
As in Paper I, we 
to neglect any transfer of angular momentum between the
disk and the star via magnetic torques.
Furthermore, we continue to ignore the effect of stellar irradiation on the structure of
the disk, although we do include its effect on the emission by the disk.

From conservation of the mass in an annulus at $\varpi$, we have
\begin{equation}
\varpi\ppbyp{\Sigma}{t}+\frac{\partial}{\partial \varpi}(\varpi \Sigma v_\varpi) =    
\varpi \ppbyp{\Sigma_\mathrm{in}}{t}-\varpi \ppbyp{\Sigma_w}{t}, 
\label{eq.mass}
\end{equation}
where $\Sigma(\varpi)$ is the surface density of the disk at $\varpi$
and $v_\varpi$ is the
radial velocity of the accretion flow (negative for inward motion).
Here we have supplemented the standard equation of mass conservation
for disks (e.g., Frank et al. 2002) with $\varpi$ times
$\partial{\Sigma}_\mathrm{in}(\varpi)/\partial t$, the mass loading rate per unit area of
the disk from the (Ulrich) inflow, and $\varpi$ times 
$\partial{\Sigma}_w(\varpi)/\partial t$,
the mass loss rate per unit area of the disk into the wind.
Integration of this relation implies the accretion rate through the disk,
$\dot{m}_\mathrm{acc}$,
varies with radius $\varpi$:
\begin{eqnarray}
\mda(\varpi) & \equiv & -2\pi \varpi\Sigma v_\varpi\nonumber\\
 &= & \pbyp{t}\left[m_d(\varpi)-m_{\rm in}(\varpi)+m_w(\varpi)\right]+\dot{m}_*,\label{eq:macc}
\end{eqnarray}
where $\partial m_d(\varpi)/\partial t$ is the rate of change of the mass of the disk
inside $\varpi$, $\partial m_w/\partial t$ and $\partial m_{\rm in}/\partial t$ are the outflow 
and
inflow rates inside $\varpi$, and $\dot{m}_*$ is the accretion rate onto
the star.  The disk extends from the stellar surface at $\varpi_*$ to
$\varpi_d$.  We use the Ulrich solution to determine the rate at which mass
flows from the envelope onto the disk. For consistency with Paper I,
we adopt the same stellar accretion rate,
$\dot{m}_*=\dot{m}_{\rm acc}(r_*)=2.4\times 10^{-4} M_\odot$~yr\e, for most of the
models we consider.

Similarly, we supplement the standard equation of angular momentum
conservation for disks (Frank et al. 2002) with terms representing the
gain and loss of angular momentum due to inflow and outflow,
respectively,
\begin{eqnarray}
\varpi\frac{\partial}{\partial t}(\Sigma \varpi^2 \Omega) & + &\frac{\partial}{\partial \varpi}
(\varpi\Sigma v_\varpi \varpi^2 \Omega)\nonumber\\
& = & \frac{1}{2\pi}\frac{\partial G_\mathrm{T}}{\partial \varpi}+
\varpi j_\mathrm{in}\ppbyp{\Sigma_\mathrm{in}}{t} 
-\varpi j_w\ppbyp{\Sigma_w}{t} , 
\label{eq.ang}
\end{eqnarray}
where $G_\mathrm{T} = 2\pi \varpi \nu \Sigma \varpi^2
(d\Omega/d\varpi)$ is the viscous torque, $j_\mathrm{in}(\varpi)$ is
the specific angular momentum of the inflow joining the disk,
$j_w(\varpi)$ is the specific angular momentum (both rotational and
magnetic) of the wind, and $\Omega$ is the angular frequency.
We have assumed that the disk extends to the surface of the protostar,
and we further assume that there is no viscous transport of angular
momentum between the disk and protostar (i.e.,
$G_\mathrm{T}=0$
at the inner edge of the disk).
Combining Equations (\ref{eq.mass}) and (\ref{eq.ang}) we
have
\begin{eqnarray} 
\frac{dj}{dt} & = & \frac{\partial j}{\partial t}+v_\varpi\frac{\partial
  j}{\partial \varpi}\nonumber\\
  & = &\frac{1}{(2\pi \varpi\Sigma)}\frac{\partial G_\mathrm{T}}{\partial \varpi}
+\frac{(j_{\rm in}-j)}{\Sigma}\ppbyp{\Sigma_{\rm
    in}}{t}+\frac{(j-j_w)}{\Sigma}\ppbyp{\Sigma_w}{t}, 
\end{eqnarray}
where $j\equiv \varpi^2\Omega$ is the specific angular momentum of the disk
gas.  We see that $j$ decreases (angular momentum is transferred
outwards) (1) due to the viscous torque, provided $\partial G_\mathrm{T}/\partial
\varpi<0$; (2) the material falls onto the disk with sub-Keplerian speed;
and/or (3) the wind leaves the disk with a super-Keplerian speed due
to magnetic torques.  Note that the deepening of the gravitational
potential well because of the growth of the star and the disk enters
into $\partial j/\partial t\propto \partial\Omega/\partial t$.

As mentioned in the previous section, we assume the disk mass always
a constant fraction of the protostellar mass,
\beq
f_d=\frac{m_d}{m_*}=\frac{\dot{m}_d(\varpi_d)}{\dot{m}_*}=1/3,
\eeq
where $\dot m_d(\vpd)$ is the rate of change of the total mass in the disk.
$\partial{\Sigma_w}/\partial t$ 
and $j_w$ are given by the wind solution 
(see Section \ref{sec:outflow}),
with the total mass loss rate to the wind also assumed to be a constant fraction of the stellar 
accretion rate,
\beq
\dot{m}_w(\varpi_d)=f_w \dot{m}_*=0.1 \dot{m}_*.
\eeq
Then the total mass loading rate from the inflow to the disk is
\beq
\dot{m}_\mathrm{in}(\varpi_d)=(1+f_w+f_d)\dot{m}_*,
\eeq
with the mass loading rate per unit area 
($\partial{\Sigma}_\mathrm{in}/\partial t$) 
and its specific angular momentum $j_\mathrm{in}$ given by the Ulrich solution.
With these boundary conditions,
from Equations (\ref{eq.mass}) and (\ref{eq.ang}),
$\Sigma(\varpi)$, 
$\partial{\Sigma}(\varpi)/\partial t$, 
and other profiles of midplane temperature,
disk scale height, 
can be solved with the $\alpha$-disk model. A detailed formulation can
be found in Appendix \ref{app:disk}.  
Here we assume $\alpha$ is a
constant throughout the disk
that is 
self-consistently determined by these boundary conditions.
In the final fiducial model of the series $\alpha=1.37$.

Although we do not explicitly use an
equation of global energy conservation in constructing our model, we
note that it can be written as
\begin{equation}
\frac{G\dot{m}_*m_*}{r_*}=L_\mathrm{acc}+L_w+L_d+\dot{E}_d,\label{eq.energy}
\end{equation}
where $G$ is Newton's gravitational constant, the l.h.s of this
equation is the total potential energy released in the accretion
process from infinity to the stellar surface, $L_\mathrm{acc}=G
\dot{m}_*m_*/(2 r_*)$ is the energy emitted as radiation when the
accreting material hits the surface of the protostar, $L_w$ is the
total mechanical energy carried away by the wind, $L_d$ is the energy
radiated from the disk
due to viscous dissipation (note that radiation from star or part of $L_\mathrm{acc}$ may
be absorbed and remitted from the disk; they are not counted in $L_d$),
and $\dot{E}_d$ is the rate of change of the
energy content (gravitational and kinetic) of the disk, which is
actually much smaller (by 2 orders of magnitude) than $L_w$ and $L_d$
in our case. For the disk wind, $L_w=1.22 L_d$ in the fiducial model
(see Section \ref{sec:outflow}), which means the disk luminosity is
now smaller by a factor of 0.45 compared with that in Paper I, since
more than half the disk energy is extracted by the wind.  The
accretion luminosity $L_\mathrm{acc}$ as it transitions from the disk
to the star ($2.45\times 10^3 L_\odot$ in the fiducial case) is
assumed to be emitted from the stellar surface homogeneously, which
indicates an effective surface temperature of the protostar of
$T_{*,\mathrm{acc}}=1.43\times10^4$~K, including the internal
luminosity ($2.82\times 10^3 L_\odot$) from the protostar.

\begin{figure}
\begin{center}
\includegraphics[width=1.\columnwidth]{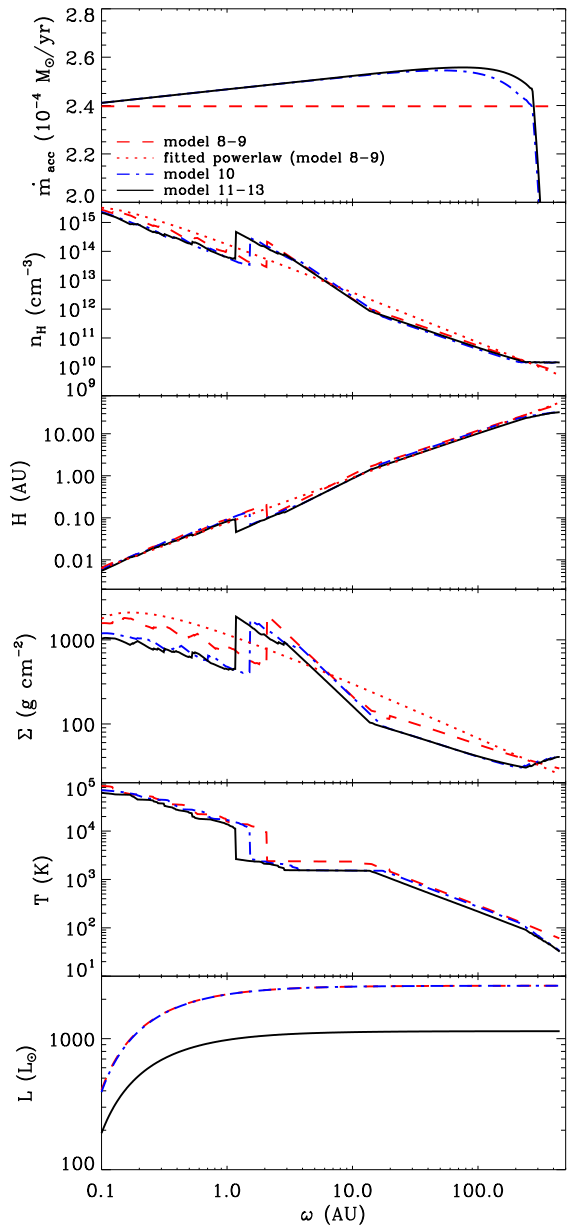}\\
\caption{The accretion rate, density, scale height, surface density,
temperature profiles, and luminosity radiated from 
a thin
disk inside cylindrical radius $\varpi$. 
$n_\mathrm{He}=0.1n_\mathrm{H}$ is assumed here. 
The dashed red curves are from a standard $\alpha$ disk 
with a constant accretion rate. The dotted red curves are 
fitted power laws, which are used in Paper I (Model 8) 
and in Model 9 in this paper. The blue curves contain the effects 
of varying accretion rate in the disk due to the inflow and the outflow,
angular momentum of the inflow,
disk growth, and a non-Keplerian potential (Model 10).
Effects of angular momentum and kinetic luminosity of the outflow are considered
in Model 11 (black lines).}
\label{fig:disk}
\end{center}
\end{figure}

\subsubsection{Thin Disk}

The disk profile is calculated numerically 
as described in Appendix \ref{app:disk}
with the opacities used in our model. Unlike in Paper I, where we 
then fit the density and scale height with power-law functions of
radius for implementation in the radiative-transfer code,
here we use the numerical solutions directly.  The accretion
luminosity is emitted from the disk, obeying the profile $4\pi
T_c^4/(3\tau)$ , where $\tau$ is the optical depth from the midplane to
the surface (see Appendix \ref{app:disk}).  
The midplane temperature, $T_c$, is calculated from the
$\alpha$-disk model. We find that it is also consistent with the
temperature from the final radiative transfer simulation, since the
heating from the star has negligible penetration to the disk midplane.

The boundary between the inflow and the outflow is
described by an \citet{Ulrich76} streamline,
which is the trajectory of an infalling particle
with an initial angular momentum but with no pressure forces:
\begin{equation}
r=\frac{\varpi_d\cos\theta_{w,\mathrm{esc}}\sin^2\theta_{w,\mathrm{esc}}}{\cos\theta_{w,\mathrm{esc}}-\cos\theta}.
\end{equation}
Here $\varpi_d\sin^2\thesc$ is the radial distance to the point at
which the trajectory crosses the disk plane ($\theta=\pi/2$); 
the disk radius is determined by trajectories in the plane ($\thesc\rightarrow \pi/2$), which gives $r\rightarrow\varpi_d$. 
Therefore, under the condition of a thin disk, the boundary between the 
wind launching region and the inflow joining region of the disk is 
$\varpi_d\sin^2\thesc$, which is 270 AU in the fiducial case.

Figure \ref{fig:disk} compares the radial profiles of the thin disk in
the various models. 
There are jumps in the profiles of density, scale height
and temperature due to the drop of the opacity from dusty conditions
at 1400~K to gas-only conditions at 1600~K and then a rapid increase
of the gas opacity with temperature to $\sim10^4$ K. The material is
piled up just before
this transition area at $T\lesssim 1600$~K. 
The red dashed curves are
numerically calculated with a standard $\alpha$ disk with a constant
accretion rate.  In Paper I (Model 8) and in Model 9 of this paper
(see Sec. \ref{sec:modelseries}), we used the fitted power laws, shown
as the red dotted lines. 
The dashed blue curves
(Model 10)
show the effects of the radial variation in the accretion rate due to
the inflow and the outflow, the effects of the angular momentum of the
inflow (but not of the outflow), the effects of a time-dependent disk,
and the effects of a non-Keplerian potential. 
The first panel shows the accretion rate through the disk (Eq. \ref{eq:macc}).
Inside 100 AU, $\mda$ increases due to the mass loss to the wind ($\partial m_w/\partial t$).
From 100 AU to 270 AU the negative $\partial m_d/\partial t$ 
(see the discussion in Appendix \ref{app:disk}) becomes dominant,
so $\mda$ starts to decrease. Outside 270 AU, $\mda$ drops rapidly due to the
inflow ($\partial m_\mathrm{in}/\partial t$).
One effect of this spatially varying accretion rate is that
the surface density of the inner disk (inside 200 AU) is less than that
of the outer disk.
The black curves show the effects of including the 
wind torque. Panels 2-5 show that this brings the dust/gas transition inward, while the last panel
shows that it reduces the viscous dissipation and the corresponding
luminosity radiated from the disk by a factor of 0.45.

\subsubsection{Effect of Disk Thickness}
\label{sec:thickness}

The finite thickness of the disk has a substantial effect on where the 
boundary between the
outflow and inflow intersects the disk surface, 
so we develop a simple approximation to allow
for this. The vertical distribution of the density in the disk ($\varpi \leq
\varpi_d$) has a Gaussian profile
determined by the temperature at the disk mid plane
(see Eq. 16 in Paper I).
In the absence of outflows, the height of the disk at $\varpi_d$ 
was defined as the height of the innermost Ulrich streamline. 
This sets a volume density at the disk surface. In Paper I, the 
disk surface at interior radii was set by having this same volume density.
In the presence of an outflow, the disk height 
in the wind-launching region
is defined by the point at which the
density in the disk equals the density at the base of the wind, which we
have somewhat arbitrarily defined as the point where the wind velocity is 1\% of
the Keplerian velocity (see Appendix B2).
For disks of finite thickness, the innermost Ulrich trajectory intersects the disk surface at
$\varpi_{w,\mathrm{max}}>\varpi_d\sin^2\thesc$.
Our method of determining the disk height in the wind-launching region becomes inconsistent with the condition that the
wind extend to the innermost Ulrich streamline
if that streamline would intersect our model disk beyond its outer edge
at $\varpi_d$. Now, we 
model the disk as being thin, so that the radiative flux in the disk is vertical, and correspondingly the outer boundary of the disk is a vertical wall at $\varpi_d$.
The problematic case--which includes the fiducial case we consider--corresponds to having
the outermost wind trajectory exit our model disk through the vertical wall instead of the disk
surface. In fact, the disk is
likely to be thinner near the outer edge, since radiation can escape radially as well as
vertically. Correspondingly, the density at the base of the wind is likely to be larger than the
value we estimate in Appendix B via a self-similar model. We adopt the simple approximation
for the problematic case that the height of the disk at the outer radius, $\zdm$, is the same as
that of the outermost wind streamline at that radius. Since we do not adjust the wind
density, this leads to a density discontinuity in the outer disk such that the disk density is
somewhat higher than the density at the base of the wind. We assume that the density at
the top of the disk remains constant as the radius decreases, until the density at the base
of the wind becomes large enough to match it; inside that point, the density at the top
of the disk is defined to be the density at the base of the wind. This approximation has
a relatively small effect: it needs to be applied only for $\varpi>400$~AU, i.e., a relatively small fraction of the disk, and it leads to a density discontinuity of only $\sim$ 20\% between the disk and the wind in this region.

\begin{figure*}
\begin{center}
\includegraphics[width=0.9\textwidth]{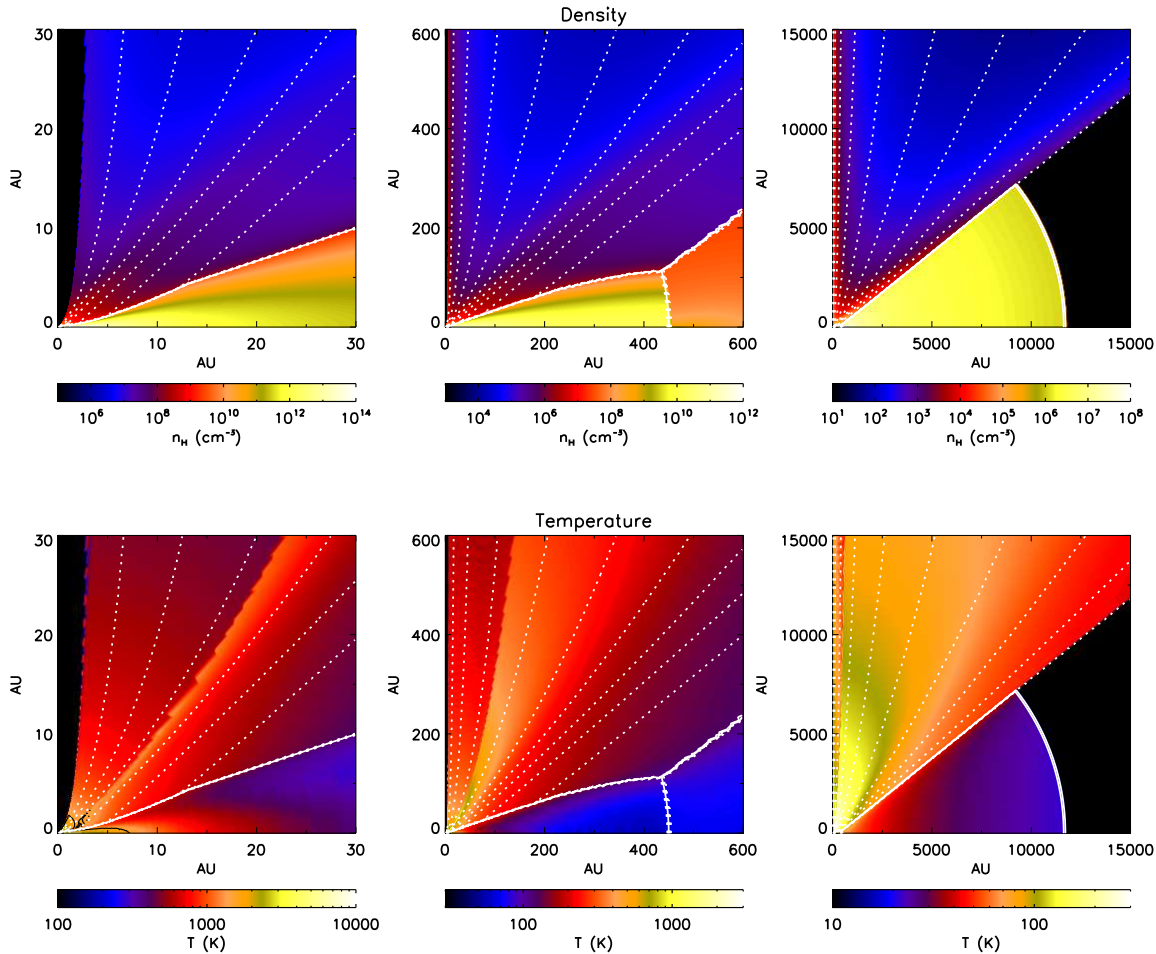}\\
\caption{The density and temperature profiles for the fiducial model (Model 13) at different scales. 
$n_\mathrm{He}=0.1n_\mathrm{H}$ is assumed.
The white contours divide the disk, the envelope and the outflow;
all three regions are shown in the middle column. The black region in the 
right column is outside the core, and is assumed to be a vacuum.
The density drops rapidly with $z$ at the base of the wind as the wind accelerates and moves outward.
The dotted lines show the streamlines of the disk wind. Each interval contains 10\% of the wind mass loss. The 
short black contour in the
lower left-hand corner of the left-most temperature plot
is the dust destruction front ($T=1600$ K). 
The dividing line between dusty and dust-free gas in that plot is apparent in the temperature jump at 10-15 AU,
this boundary starts at about 2 AU.}
\label{fig:trho}
\end{center}
\end{figure*}

\subsection{Outflows}
\label{sec:outflow}

Bipolar outflows are ubiquitous around protostars
(e.g. \citealt[]{KP00}). These outflows (or winds: we will use these
two terms interchangeably in this paper) are believed to be driven by
the magnetic fields threading the rotating disk (or the star) and
powered by the accretion. Current wind models can be categorized into
two classes: disk winds, in which the wind flows along the magnetic
field lines threading the disk (\citealt[]{KP00}), and X-winds, in
which the wind is launched from a narrow interaction region between
the stellar magnetic field and the inner edge of a truncated accretion
disk (\citealt[]{Shu00}). Although the properties of these two winds
in the near-disk region are quite different, their momentum
distributions become similar at regions far from the disk, with
$p_w\propto (r\sin\theta_w)^{-2}$ (\citealt[]{MM99}). In Paper I, this
property, together with an assumed star formation efficiency from the
core, was used to determine the opening angle of the outflow cavities,
which however were left empty. Here, we now include a detailed density
distribution for such outflows.

In this paper we consider only disk winds.  We base our analysis on
the wind calculated by \citet[we hereafter refer to this as the BP
  wind]{BP82} and refined by \citet[]{CL94}, but we alter this so that
the outermost streamline of the wind, at cylindrical radius $\vpm(z)$,
coincides with the innermost streamline of the accretion flow.  In the
Ulrich solution for the accretion flow, this streamline intersects the
disk midplane at $\varpi=\varpi_d\sin^2\theta_{w,\mathrm{esc}}$.  We
assume that there is a central cavity in the wind due possibly to the
poloidal field of the protostar, as in the X-wind model
(\citealt[]{Shu95}), or to a line-driven wind from the protostar; the
radius of this cavity is $\vpc(z)$. For X-winds, one can show that
$\vpc(z)$ is qualitatively similar to a BP streamline; since we are
considering a BP-like wind here, we assume that $\vpc(z)$ does in fact
have the shape of a BP streamline. Wind streamlines begin at
a cylindrical radius $\vpo$ and at a height $z_d(\vpo)$.
We assume that the innermost wind streamline emanates from the surface of
the star, $\vpco=r_*$.  BP winds have a density at the base of the
outflow that varies as $\vp^{-3/2}$ and a velocity that scales as the
Keplerian velocity, $v_{\rm K}\propto \varpi^{-1/2}$. As a result, the mass
loss scales logarithmically with the radius of the outermost
streamline, 
\beq 
\mdw=4\pi\int_{r_*}^{{\vpmo}}\vpo\rho v_z d\vpo \propto \ln(\vpmo/\vpco),
\eeq 
where we have assumed that the disk is thin, so that the mass flux is
determined by the $z$-component of the velocity.  The density in the
outflow at a height $z'$ above the disk surface is given in Appendix
\ref{app:outflow}, in terms of the cylindrical radius of the
footpoint, $\vpo$.

As discussed in the last section, the outflow extracts angular
momentum and accretion energy from the disk. The specific angular
momentum of a self-similar disk wind is (e.g. \citealt[]{KP00})
\begin{equation}
j_w=\Omega \varpi_A^2,
\end{equation}
where $\varpi_A$ is the cylindrical radius of the Alfv\'en surface
(where the wind becomes super-Alfv\'enic). The ratio between the
Alfv\'en radius and the footpoint radius of a streamline,
$\varpi_A/\varpi_0$, can be considered as a constant and has typical
value of $\sim 3$ (e.g. \citealt[]{KP00}). For the BP wind we
approximate here (see Appendix \ref{app:outflow}),
$\varpi_A/\varpi_0=5.48$. For such a wind, the specific energy of the
wind, $e_w$, is given by (\citealt[]{BP82})
\begin{equation}
\frac{e_w}{\Omega^2 \varpi_0^2}=\frac{j_w}{\Omega \varpi_0^2}-\frac{3}{2}=\left(\frac{\varpi_A}
{\varpi_0}\right)^2-\frac{3}{2}
=28.5.
\end{equation}

\subsection{Dust and Gas Opacities}

Dust shapes the SED of the protostar by transforming visible and UV
radiation into the infrared. In Paper I, the dust was assumed to be
completely destroyed by sublimation at a temperature
$T_\mathrm{sub}=1600$ K, which made the transition between dust and
gas opacities relatively sharp. In order to improve the temperature
convergence we make a smoother transition between the dust and gas
opacities by linearly reducing the fraction of the dust content from
100\% (gas-to-dust ratio 100) at 1400 K to 0\% (no dust content) at
1600 K, approximating the results of \citet[]{Semenov03}. 
We identify 1600 K as a reference sublimation temperature.

We assume that the wind is dusty if it is launched outside the radius
at which dust sublimates, neglecting other dust destruction processes
such as shocks.  Even though the temperature in the the wind launched
from the inner dust-free disk soon drops below the dust sublimation
temperature, this part of the wind is still likely to remain dust-free
since dust grains are not expected to form in such a short time at
these densities. We therefore define the boundary between the
dust-free and dusty outflow by a streamline starting from the disk
surface where the surface temperature is at the sublimation
temperature (1600 K). 
In our fiducial model, this radius is $\sim$2~AU.
Figure \ref{fig:trho} shows the streamlines (white dotted curves) and
the density profile (color scale in the upper panels). Each interval
of the streamline contains 10\% of $\dot{m}_w$.  The dusty and
dust-free parts of the wind can be recognized in the temperature
profile. About 66\% of the wind is dusty.
Note because of the finite disk thickness, 
the outermost wind streamline emerges from the disk surface at $\varpi_d$,
as discussed in Section \ref{sec:thickness}.
On the other hand, we used the thin-disk equations in
calculating the dynamics and vertical profile of the disk, 
and in that approximation the wind emerges from the midplane
at $\varpi_d \sin^2\theta_{w,\mathrm{sec}}$ (see Fig. \ref{fig:disk}).

\begin{table*}
\begin{center}
\caption{Properties of features included in the model series.\label{table}}
\ \\
\begin{tabular}{c|c|c|c}
\hline\hline
\multicolumn{4}{l}{$m_*=8 M_\odot$, $L_*=2.8\times 10^3 L_\odot$, $M_\mathrm{c}=60 M_\odot$, $R_\mathrm{c}=0.06$ pc, $\theta_{w,\mathrm{esc}}=51^\circ$, $\varpi_d=4.5\times10^2$ AU}\\
\hline\hline
\bf Models & \bf Opacity & \bf Disk & \bf Outflow \\
\hline
\multirow{2}{*}{8} & gas ($T>1600$ K) & standard $\alpha$ disk model & \multirow{2}{*}{empty}\\
& dust ($T<1600$ K)& with constant accretion rate & \\
\hline
\multirow{2}{*}{9} & smoother transition & \multirow{2}{*}{as above} & \multirow{2}{*}{empty} \\
& 
(1400 K $<T<$ 1600 K)
& & \\
\hline
& & effects of inflow / massive growing disk on & \\
10 & as above & accretion rate / angular momentum distribution; & empty \\
& & effects of outflow on accretion rate & \\
\hline
\multirow{2}{*}{11} & \multirow{2}{*}{as above}& including effects of outflow in extracting angular & \multirow{2}{*}{empty} \\
& & momentum and accretion energy from disk & \\
\hline
\multirow{2}{*}{12} & \multirow{2}{*}{as above}& \multirow{2}{*}{as above} & \multirow{2}{*}{dust free} \\
& & & \\
\hline
\multirow{2}{*}{13} & \multirow{2}{*}{as above}& \multirow{2}{*}{as above} & \multirow{2}{*}{dust + gas} \\
 & & & \\
\hline\hline
\end{tabular}
\end{center}
\end{table*}

\begin{table*}
\begin{center}
\caption{Parameters of models comparing cores with different surface density. \label{table2}}
\ \\
\begin{tabular}{l|c|c|c}
\hline
& Model 13l & Model 13 & Model 13h\\
\hline
initial core mass $M_\mathrm{c}$ ($M_\odot$) & 60 & 60 & 60\\
\hline
mean surface density $\Sigma_\mathrm{cl}$ (g/cm$^2$) & 0.316 & 1 & 3.16\\
\hline
core radius $R_\mathrm{c}$ (pc) & 0.10 & 0.057 & 0.032\\
\hline
outer radius of disk $\varpi_d$ (AU) &  801.4 & 449.4 & 253.4\\
\hline
stellar accretion rate $\dot{m}_*$ ($M_\odot$/yr) &  $1.035\times 10^{-4}$ & $2.398\times 10^{-4}$ & $5.667\times 10^{-4}$ \\
\hline
stellar radius $r_*$ ($R_\odot$) & 11.3 & 12.0 & 5.93\\
\hline
isolated
stellar surface temperature $T_*$ (K) & $1.25\times 10^4$ & $1.22\times 10^4$ & $1.74\times 10^4$\\
\hline
isolated stellar + accretion
 temperature $T_{*,\mathrm{acc}}$ (K) & $1.37\times 10^4$ & $1.42\times 10^4$ & $2.62\times 10^4$\\
\hline
isolated
stellar luminosity $L_*$ ($L_\odot$) & $2.82 \times 10^3$ & $2.81\times 10^3$ & $2.84\times 10^3$\\
\hline
stellar accretion
luminosity $L_\mathrm{acc}$ ( $L_\odot$) & $1.13\times 10^3$ & $2.45\times 10^3$ & $1.18\times 10^4$ \\
\hline
disk luminosity $L_d$ ( $L_\odot$) & $5.55\times 10^2$ & $1.10\times 10^3$ & $5.44\times 10^3$ \\
\hline
total luminosity $L_\mathrm{bol}$ ($L_\odot$) & $4.51\times 10^3$ & $6.46\times 10^3$ & $2.01\times 10^4$\\
\hline
\end{tabular}
\end{center}
\end{table*}

We divide the dusty part of the simulation into four different
regions: (1) the envelope, (2) the low density regions of the disk
($n_{\mathrm{H}}< 2\times10^{10}\mathrm{cm}^{-3}$), (3) the high
density regions of the disk, and (4) the dusty part of the
outflow. Opacity models for the first three regions were described in
Paper I. The opacity model for the outflow has smaller grains than
that in the envelope and has no ice mantles (\citealt[]{KMH94}). These
dust opacity models are default options in \citeapos{Whitney03a}
radiation transfer code.

Gas opacities are used in the dust-free, hot inner disk and in the
outflow launched from it. The mean gas opacity at $\sim$ 1500 K is
about 3 orders of magnitude lower than the opacity of the dust at
$\sim 1400$~K, but becomes comparable or even higher when the
temperature reaches $\sim 10^4$ K,
where the gas becomes collisionally ionized (note that
we are not including photoionization from the protostellar radiation
field). For temperatures higher than $\sim 3000$ K, we adopt gas
opacities from the OP Project (e.g. \citealt[]{Seaton05}). For regions
with $T< 3000$ K, the opacity model is from \citet[]{AF94}.  However,
the opacity given by \citet[]{AF94} has a minimum at $2000$ K and
increases when $T$ drops below that because of molecule and dust
formation. 
Since we expect minimal dust formation to occur in the rapidly
expanding outflow, in its dust-free region where $T<2000$~K we adopt
the opacity at $2000$~K.
More details about these gas opacities are given in Paper I.

\subsection{Simulations}
\label{sec:simulations}

We use the latest version of the Monte Carlo radiation transfer code
by \citet[2012 in prep.]{Whitney03b} to perform
our calculations. To calculate the equilibrium temperature, we use the
algorithm by \citet[]{Lucy99}, which is implemented in the new version
of the code. It sums the pathlengths of all the photons passing
through the cell rather than counting only those that are absorbed
(e.g., \citealt[]{BW01}, used in Paper I), and it thus can quickly
reach temperature convergence with fewer photon packets. Starting from
a uniform and cold ($T=$ 0.1 K) state, the temperature profile becomes
stable within $\sim 5$ iterations with a number of photon packets
equal to the number of grid cells ($\sim 1.5\times 10^6$). There are
still some oscillations at the transition regions between gas-only and
dusty opacities because of the U-shaped opacity curve as a function of
temperature, but they occur mostly
in small, localized regions close to the disk midplane. The
temperature profile on the disk surface and in the outflow is quite
stable. We iterate 20 times, average the temperature profiles of the
last 10 iterations, and then use this averaged temperature profile as
the equilibrium temperature for the final iteration.

Corrections made by adiabatic (expansion) cooling and (compressional)
heating in the outflow and in the accretion flow are also considered.
We assume that the temperature of the gas varies only slowly, so that
it is approximately in thermal equilibrium.  The thermal energy
equation for cells in these regions is then
\begin{eqnarray}
4\sigma T^4 \kappa_P(T)\rho + P_\mathrm{gas}
\nabla \cdot \mathbf{v} & + & \nabla \cdot (u\mathbf{v})\nonumber\\
&= &\left(\frac{L}{N}\right)\left(\frac{\rho\sum{\kappa_\nu l}}{V}\right),
\label{eq:thermal}
\end{eqnarray}
where $\kappa_P$ is the Planck mean opacity, $\mathbf{v}$ is the
velocity field, $P_\mathrm{gas}$ 
is the pressure of the gas, $u$ is the internal energy
of the gas, 
$L$ is the total luminosity, $N$ is the total number of photon
packets, $V$ is the volume of the cell, and $\rho\sum{\kappa_\nu l}$
sums all the optical depths that photon packets have traveled across
the cell (\citealt[]{Lucy99}). From left to right, the four terms in
the equation are the emission, the adiabatic heating or cooling,
advection, and the absorption respectively.
Adiabatic cooling becomes important in
the outer region ($r\sim R_\mathrm{c}$) of the outflow, where the
temperature drops to $< 100$ K, since the adiabatic cooling rate is
proportional to $T$ whereas the thermal emission term is proportional to $T^4$. 
In these regions adiabatic expansion reduces the temperature by a
factor of 2 to 3.
Adiabatic heating in the infalling envelope is unimportant (the emission
term is $\sim 10^3 - 10^4$ larger than the adiabatic heating term). 
Note that Equation (\ref{eq:thermal}) is valid
for both the gas and the dust only in regions where they 
are well coupled, which happens for densities
$n_\mathrm{H}>10^5~\mathrm{cm}^{-3}$ in dense molecular clouds
(\citealt[]{DN97}). This density threshold is at a radius $\sim$ 1000
AU in the outflow. Beyond this radius, the calculated value of the
dust temperature should remain valid, although the gas will become
colder. The temperature profiles are shown in Figure
\ref{fig:trho}. 
The low-temperature region in the outflow that extends to a 
cylindrical radius of $\sim$ 100 AU at height of 600 AU is
due to the lower opacity of the dust-free gas. The black region along
the axis is the inner cavity due to the stellar magnetic field or the line-driven stellar wind.

A 1000 $\times$ 1500 grid is used to resolve the azimuthally-symmetric
$r\times \theta$ space. In $r$ space, $\sim$ 750 cells are used to
resolve the disk with a finer grid at smaller $r$ ($\sim$ 400 cells
are used to cover the region inside 5 AU). In $\theta$ space, an
especially fine grid is made to resolve the midplane as well as the
inner and the outer boundary of the outflow. About 700 cells are used
between $20^\circ$ above and below the disk midplane. About 100 cells
are used within $\theta=5^\circ$, and $\sim$ 300 cells between
$\theta=50^\circ$ and $53^\circ$ (the opening angle of the outflow
cavity is 51$^\circ$). For each run, a high S/N SED and images at
different wavelengths and instrument bands ({\it Spitzer} IRAC and
MIPS, 2MASS J, H, K bands, {\it GTC} CanariCam, {\it Herschel} PACS and
SPIRE, and {\it SOFIA} FORCAST) are produced for viewing angles of
$30^\circ$ and $60^\circ$. For most of the models shown here, $10^8$
photon packets are used to obtain SEDs and $5\times10^8$ photon packets are
used to produce images.

\subsection{Model Series}
\label{sec:modelseries}

Following the same methodology of Paper I, we develop a series of
models to show the effects of the improvements discussed above. We
start this series with Model 8, the last model in Paper I, but now
calculated with the new version of the code. In Model 9, we include
the smoother transition between the gas opacities and dust opacities
in the disk. In Model 10, we improve the disk by adding the effects of
varying accretion rate in the disk due to the inflow and the outflow,
angular momentum of the inflow, disk growth, and a non-Keplerian
potential.
The effect of the
outflow in extracting angular momentum 
and accretion energy
from the disk is included in Model
11. In Model 12, we apply the wind solution to the outflow cavity but
assume it to be dust-free (i.e., only gas opacities in the outflow
cavity). We also start to consider the flow terms in the energy
balance equation from here. Model 13 is the fiducial model, which has
all the elements discussed above: a smooth transition between dusty
and dust-free gas, an accretion disk structure that includes the
effects of time dependence and outflows, and an outflow that is dusty
when launched outside the sublimation radius. 
The density and temperature structure of Model 13 is shown in Figure (\ref{fig:trho}).
The properties of each
model are summarized in Table \ref{table}.

We also construct two variants of Model 13, one with
$\Sigma_\mathrm{cl}=0.316$ g cm$^{-2}$ (Model 13l) and another one
with $\Sigma_\mathrm{cl}=3.16$ g cm$^{-2}$ (Model 13h). The size of
the core scales as $\scl^{-1/2}$, the density as $\scl^{3/2}$, the
pressure as $\scl^2$, and the accretion rate as $\scl^{3/4}$
(MT03). The parameters we use here correspond to a change in the
clump surface density by a factor of 
0.5 dex and the surface pressure by a factor of 
1 dex.
Table \ref{table2} compares the parameters of Model
13l, 13, and 13h.

\section{Results and Discussion}
\label{sec:result}

\subsection{SEDs}
\label{sec:sed}

\begin{figure}
\begin{center}
\includegraphics[width=\columnwidth]{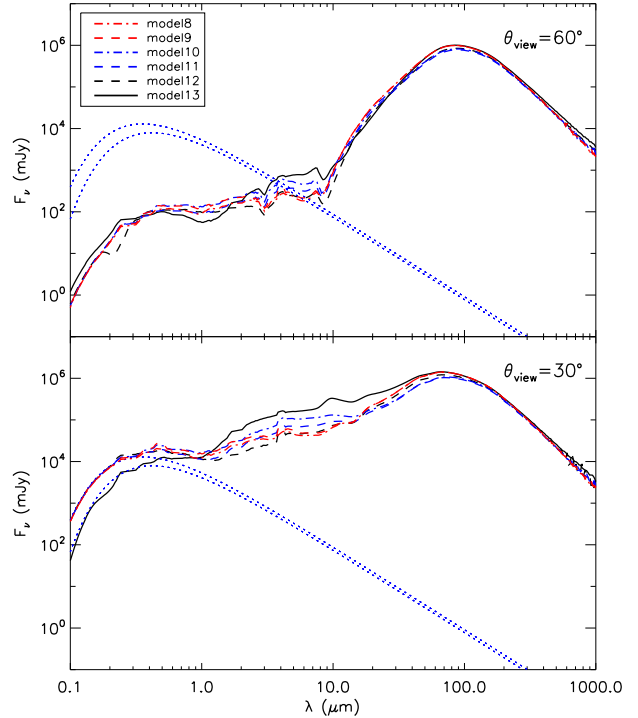}\\
\caption{SEDs for the model series, assuming a distance of 1 kpc. Two typical inclinations of $60^\circ$ and $30^\circ$ are shown. The lower dotted blue curve is the input stellar spectrum (black-body), and the upper dotted one is the total luminosity of the star,
including both the internal luminosity and the energy released by the
gas as it accretes from the disk to the star. We assume that this
radiation is emitted as a black-body.}
\label{fig:sed_series}
\end{center}
\end{figure}

\begin{figure}
\begin{center}
\includegraphics[width=\columnwidth]{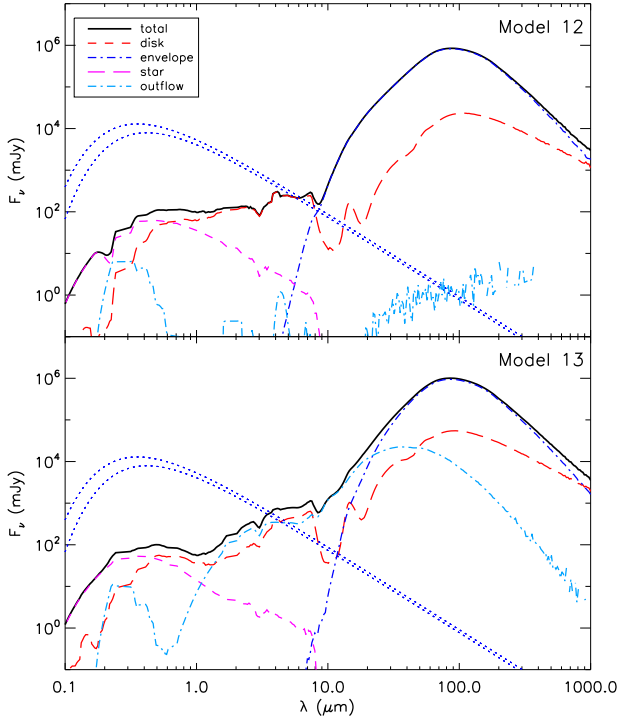}\\
\caption{SEDs at an inclination of $60^\circ$ and a distance of 1 kpc, comparing Model 12 and Model 13. The total flux and the flux from each component are shown. The two dotted blue curves are the input stellar spectra associated with the internal luminosity and the 
internal 
plus accretion luminosity, emitted as a black-body in each case.}
\label{fig:sed_compo}
\end{center}
\end{figure}

\begin{figure}
\begin{center}
\includegraphics[width=\columnwidth]{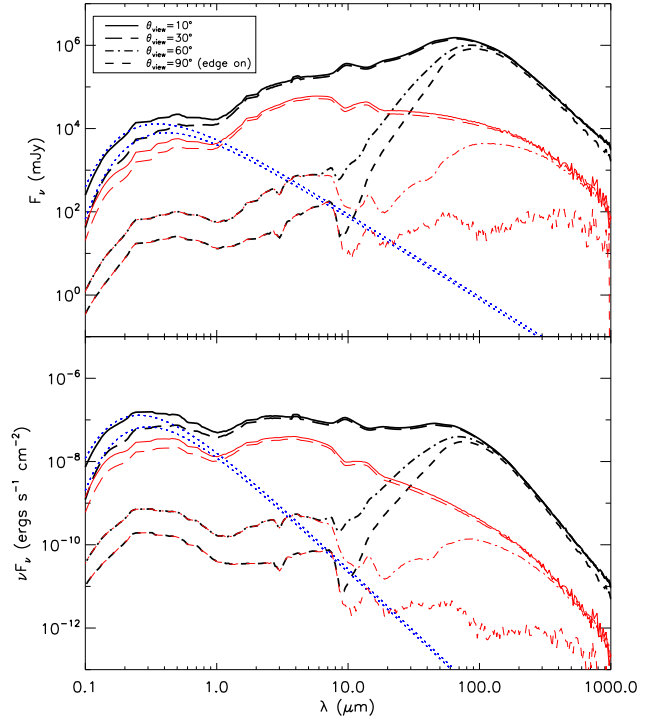}\\
\caption{SEDs of the fiducial model (Model 13) at four
typical inclinations, assuming a distance of 1 kpc ($F_\nu$ in the upper panel and $\nu F_\nu$ in the lower panel). The thicker curves are total fluxes and the thinner red curves show only the scattered light. The two dotted blue curves are the input stellar spectra associated with the internal luminosity and the internal plus accretion luminosity, emitted as a black-body in each case.}
\label{fig:sed_fiducial}
\end{center}
\end{figure}

Figure \ref{fig:sed_series} shows the SEDs of the models at
inclination angles between the line of sight and the rotational axis
of $60^\circ$ and $30^\circ$. At an inclination of $60^\circ$, the
line of sight passes through the dense envelope, which obscures the
protostar and the disk. At an inclination of $30^\circ$, the line of
sight passes through the outflow cavity; because the density in the
cavity is lower than in the envelope, and because only part of the
outflow is dusty, we can see much more deeply into the cavity than
into the envelope. Some important features in the SEDs include the
water ice feature at 3 $\mu$m and the silicate 
feature at 10 $\mu$m.  As discussed in Paper I, the original output SEDs show
some significant emission features such as the H$\alpha$ and Paschen
$\alpha$ lines, but we subtract these features and smooth the SEDs at
wavelengths $<2\mu$m, since the line profiles may not be exactly
correct given the spectral resolution of the radiative transfer
simulation and since this line emission is not the focus of our study.

The differences of the SEDs among Models 8-12 are relatively small. The
angle-dependent SEDs of Model 8 are the same as in Paper I, showing
that the new method of calculating the temperature produces the same
results as the method used in the previous paper. Model 9 has a
smoother transition between the gas and dust opacities, which is both
physically more realistic and achieves better temperature convergence.
Model 10 and Model 11 update the disk profile by including the effects
of the inflow, the massive disk, and the outflow. Compared with Models
8 and 9, they have less dense but thicker inner disks,
so that starlight can heat the inner parts of the disk 
more easily, producing warmer disk surfaces, while the base of the
envelope is more shielded and becomes cooler, thereby making the near-
and mid-IR emission higher and the far-IR peaks lower. In Model 11,
half of the accretion energy drives the outflow, reducing the emission
compared to that in Model 10, particularly between 1 $\mu$m and 10
$\mu$m. In Model 12, gas opacities are included in the outflow
cavity. However, most of the outflow is very cool so that the gas
opacity is low, making the SEDs very similar to those of Model 11.

The dust in the outflow cavity affects the SED more significantly.  As
shown by the SED of Model 13 at a viewing angle of 30$^\circ$ in 
Figure \ref{fig:sed_series}, the
emission at short wavelengths ($\lambda\la 1 \mu$m) 
is now suppressed. The reprocessed emission comes out as extra near-
and mid-infrared radiation.  At larger inclinations, the fluxes at
short wavelengths become higher than in the previous models due to
scattering by dust grains.  Figure \ref{fig:sed_compo} compares the
SEDs of Models 12 and 13 at an inclination of $60^\circ$ in more
detail. Flux components of different origins are also shown.  The
existence of dust in the outflow cavity makes the disk warmer,
producing higher mid- and far-IR fluxes.  The outflow also becomes
hotter and emits much more infrared radiation, especially around 10 to
20 $\mu$m, where the outflow emission becomes dominant.  Depending on
the sensitivity of the observations, some of the differences among
these models may not be detectable, such as from Model 8 -
12. However, the dust in the disk wind (Model 13) can produce 
factors of several
differences to the SED in near- and mid-IR, which can be
constrained by the observations.
As we show below, dust in the outflow cavity also has significant
effects on the infrared images,
which are likely to be more useful for observationally testing
these components of the theoretical model.

\begin{figure*}
\begin{center}
\includegraphics[width=\textwidth]{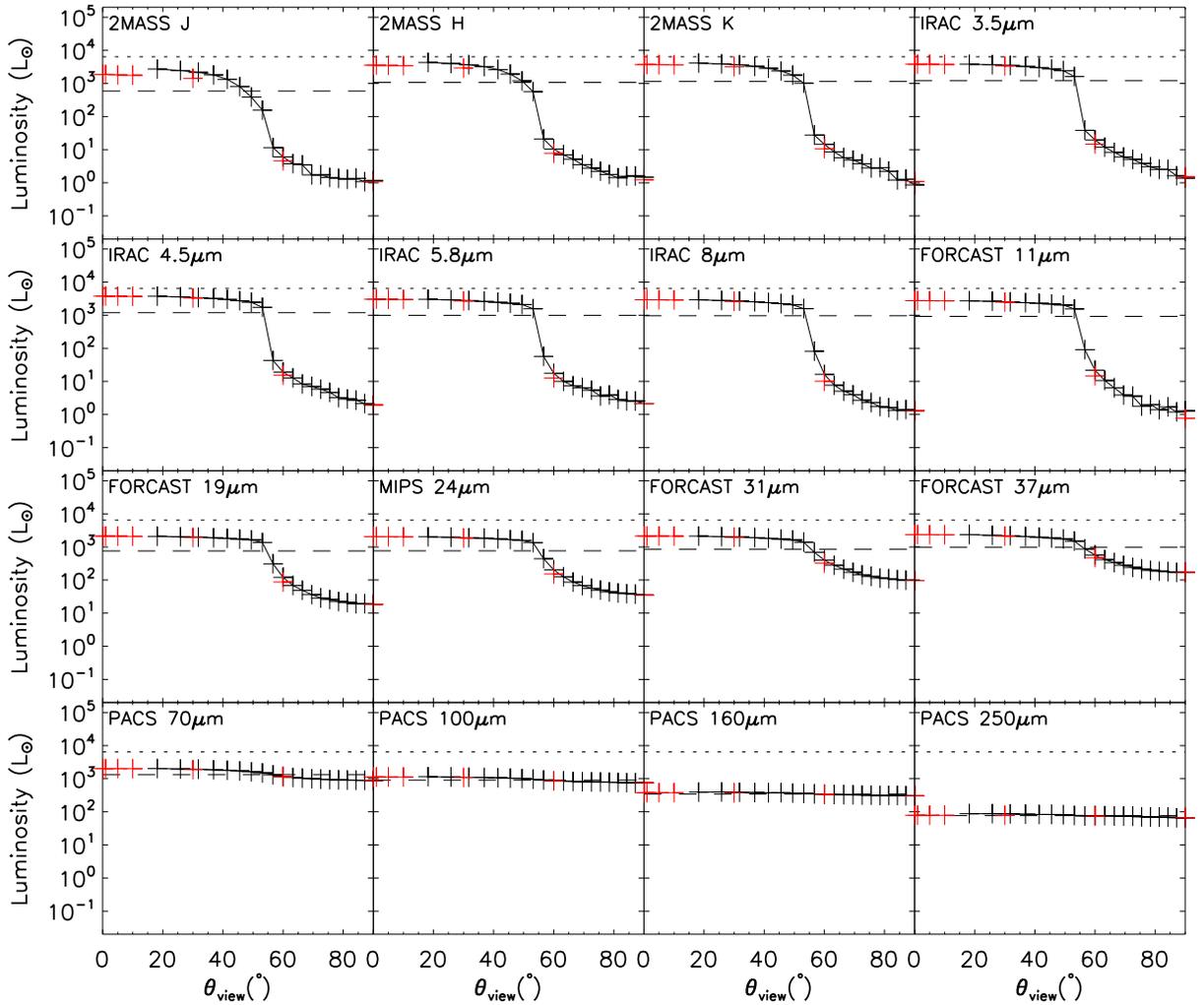}\\
\caption{Inferred IR luminosities at different viewing angles and at different
wavelengths (plus symbols and curves), compared with the true IR
luminosities at these wavelengths (dashed lines) and the bolometric
luminosities (dotted lines). The black plus symbols are at 20 viewing
angles with $\cos\theta_\mathrm{view}$ evenly distributed.
The red plus symbols are at viewing angles
of 0.1$^\circ$, 1$^\circ$, 5$^\circ$, 10$^\circ$, 30$^\circ$, 60$^\circ$, and 90$^\circ$.
 The data shown with red plus symbols
have a higher S/N than those in black.}
\label{fig:flashlight}
\end{center}
\end{figure*}

\begin{figure}[here]
\begin{center}
\includegraphics[width=\columnwidth]{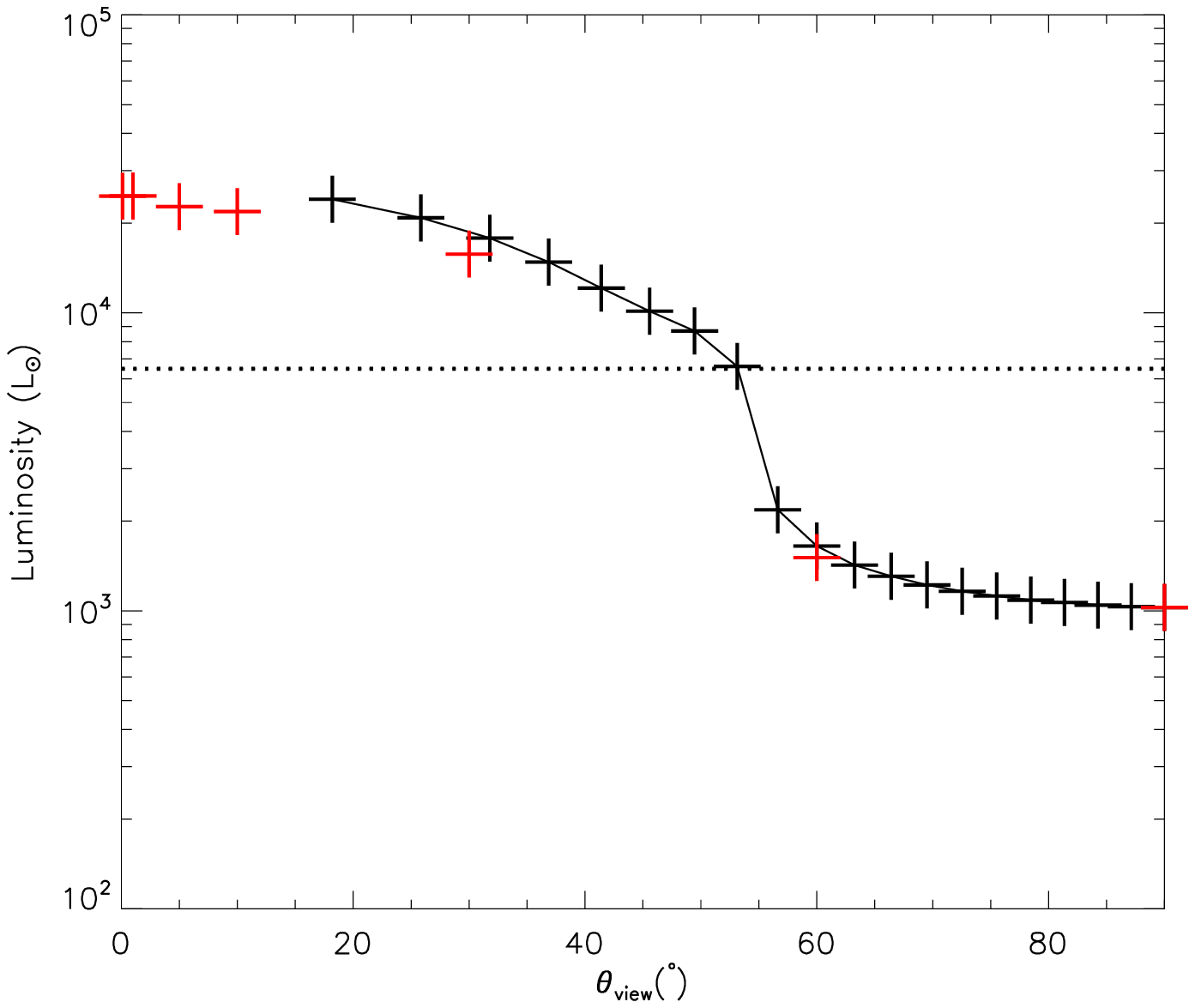}\\
\caption{Inferred bolometric luminosity at different viewing angles
(plus symbols and curves), compared
with the true bolometric luminosity (the dotted line). The meaning of
the black and red symbols are same as Fig. (\ref{fig:flashlight}).}
\label{fig:flashlight_bol}
\end{center}
\end{figure}

The SEDs of the fiducial model (Model 13) at four inclinations are
shown in Figure \ref{fig:sed_fiducial}.\footnote{We note that in our highly idealized model, in which the axis of the
outflow is exactly straight and the BP streamlines persist to large
distances, it is possible to see the 
protostar 
through the dust-free part of 
the outflow
only for viewing angles that are very small. For Model 13, in which
the dusty outflow commences at $\vpo=2$~AU and the core size is
$R_c=1.2\times 10^4$~AU, the central dust-free cavity has a size of
about 600 AU at $R_c$, corresponding to an opening angle of about
$3^\circ$.
At greater distances from the protostar, this angle becomes even
smaller.  In fact, however, protostellar outflows are not observed to
remain exactly straight to large distances, so it is quite possible
that there is no inclination at which the 
protostar is 
completely
unobscured by outflow dust. However, in the fiducial model considered here, the amount of obscuration by outflow dust 
is negligible for moderate angles of inclination ($\sim 10^\circ$).
} Here we extend the wind to $5R_\mathrm{c}$. If the wind
extended to $10R_\mathrm{c}$, $A_V$ would increase by $\sim$ 1\% at an
inclination of 10$^\circ$, and $\sim$ 0.1\% at 30$^\circ$.  Since the
density of the wind drops as $\rho\propto\varpi^{-2}\sim r^{-2}$ (see
Appendix \ref{app:outflow}), we conclude that the SEDs are not
affected by truncating the wind at $5R_\mathrm{c}$.
As a result of dust in the outflow, the optical emission is suppressed
by a factor of 2--3
when the line of sight passes through the outflow cavity at a viewing
angle of 30$^\circ$). However, at a viewing angle of 10$^\circ$, the extinction is negligible and
the SED 
closely follows the stellar spectrum at short wavelengths.
The SEDs
(in $\nu F_\nu$)
at these viewing angles are relatively flat from the near-IR to about
60 $\mu$m, which may be used as an identifier of near face-on massive
protostars.
The red curves show the SEDs in scattered light. If the line of sight
passes through the envelope, all the light at wavelengths shorter than
$\sim 10 \mu$m has been scattered. 
Direct
emission dominates for models with lines of sight passing through the
outflow cavity, even at short wavelengths.

As the extinction varies with inclination of the viewing angle, the IR
(or bolometric) luminosity inferred from the observed flux varies,
which is known as the ``flashlight effect'' (
\citealt[]{Nakano95}, \citealt[]{YB99}).  Figure
\ref{fig:flashlight} shows how the inferred IR luminosities from the
near-IR to the far-IR change with viewing angle. In the near-IR, the
flux for a face-on view can be higher than that for an edge-on view by
3 orders of magnitude,
and for wavelengths $\la 40 \;\mu$m, it can be larger than the value averaged over
the sky by a factor $\sim 3$.
This contrast decreases as the wavelength
increases; at 100 $\mu$m and longer wavelengths, it becomes very
small.  When the line of sight passes through the outflow cavity
close to face-on, the inferred IR luminosities (from the observed $\nu F_\nu$) in the
(1 - 8)~$\mu$m band give a good estimate of
the total bolometric luminosity,
being low by a factor of $\sim$2.
Figure \ref{fig:flashlight_bol} shows how the bolometric luminosities 
inferred from the SED depend on the viewing angle and how they compare with the
true bolometric 
luminosities. When looking down the outflow cavity, the bolometric 
luminosity inferred from the SED is higher than
the true value by a factor of $\sim$ 3
The inferred luminosity
gradually decreases with increasing inclination angle and drops rapidly when the
angle is large enough that the line of sight passes through the envelope. 
At high inclinations, the inferred bolometric luminosity is lower than its true value by 
a factor of $\sim$ 6.

\begin{figure}[here]
\begin{center}
\includegraphics[width=\columnwidth]{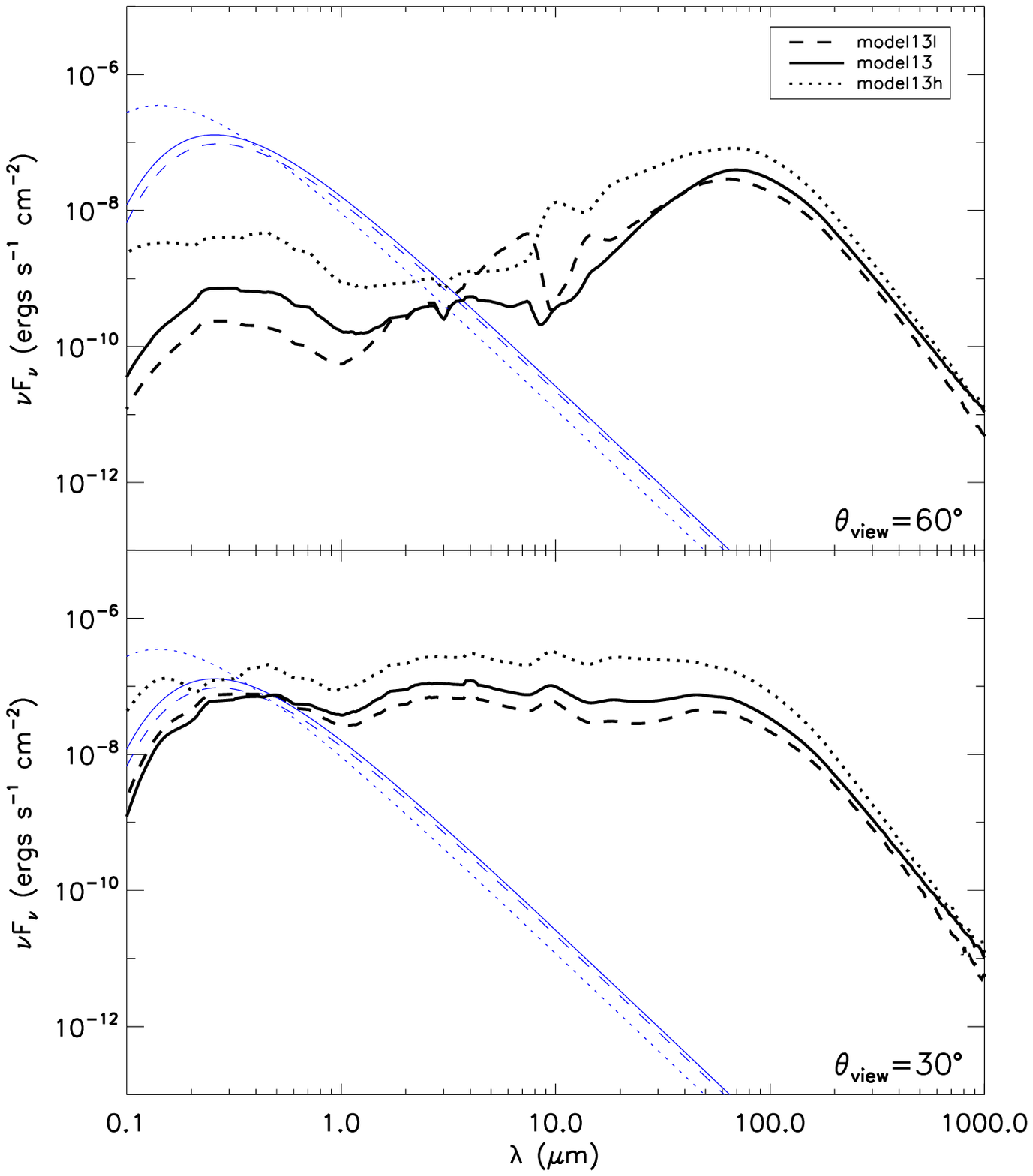}\\
\caption{SEDs of Model 13l ($\Sigma_\mathrm{cl}=0.316$ g cm$^{-2}$), Model 13 ($\Sigma_\mathrm{cl}=1$ g cm$^{-2}$), and Model 13h ($\Sigma_\mathrm{cl}=3.16$ g cm$^{-2}$) at inclinations of 60$^\circ$ (upper panel) and 30$^\circ$ (lower panel), assuming a distance of 1 kpc. The black curves are the total emission, and the blue curves are the input spectra of the protostar 
(internal + accretion).
As discussed in the text, the protostar in Model 13h has a smaller radius, and therefore a larger accretion luminosity, than those in Models 13 and 13l.}
\label{fig:sed_sigma}
\end{center}
\end{figure}

Figure \ref{fig:sed_sigma} compares the SEDs of Model 13, Model 13l
and Model 13h, which have cores with the same mass but which are
embedded in clumps of different surface densities,
$\Sigma_\mathrm{cl}$. As noted above, the density in the core scales
as $\Sigma_\mathrm{cl}^{3/2}$, which has two effects on the SED:
First, since the accretion rate scales inversely with the free-fall
time, which varies as $\rho^{-1/2}$, the accretion luminosity, which
scales with the accretion rate, varies as $\Sigma_\mathrm{cl}^{3/4}$;
this affects the SED at all wavelengths. 
Second, since the surface density of the core scales linearly with
that of the clump in which it is embedded, the extinction 
due to
the core is proportional to $\Sigma_\mathrm{cl}$; this affects the SED
mainly at mid-IR and shorter wavelengths. 
Note, that we are not including the additional extinction that would
result from the clump itself.
So, as shown in the SEDs, Model 13h generally has higher fluxes at all
wavelengths than Model 13 and Model 13l, due to the much higher
luminosity.  At lower inclination, the higher extinction due to the
outflow makes the FUV fluxes in Models 13 lower than those in Model
13l (although these fluxes will in general be difficult to observe due
to additional foreground extinction from dust in the clump and the
diffuse Galactic ISM).  At higher inclination, the lower extinction of
the core makes the mid-IR flux of Model 13l higher than that of Model
13. In this case, the mid-IR flux from the disk in Model 13l is higher
and less buried by the flux from the envelope, producing a more
prominent 10$\mu$m silicate absorption feature than in Model 13.  This
again implies that the depth of the silicate feature depends not only
on the optical depth but also on the ratio of the mid-IR emission from
the disk to that from the envelope.

An additional factor is the dependence of the protostellar evolution
(especially size of the protostar) on the accretion rate. The
protostar in Model 13h is much smaller (by a factor of about 0.5) in radius
than the other models: it has not yet reached the luminosity wave
stage that swells the star (MT03). Thus, its accretion luminosity is
significantly higher than that expected from just its enhanced
accretion rate. The dependence of these results on the use a
simplified one zone protostellar evolution model (MT03) compared to a
more detailed treatment, will be examined in a future paper.

\subsection{Images}
\label{sec:image}

\begin{figure*}
\begin{center}
\includegraphics[width=0.97\textwidth]{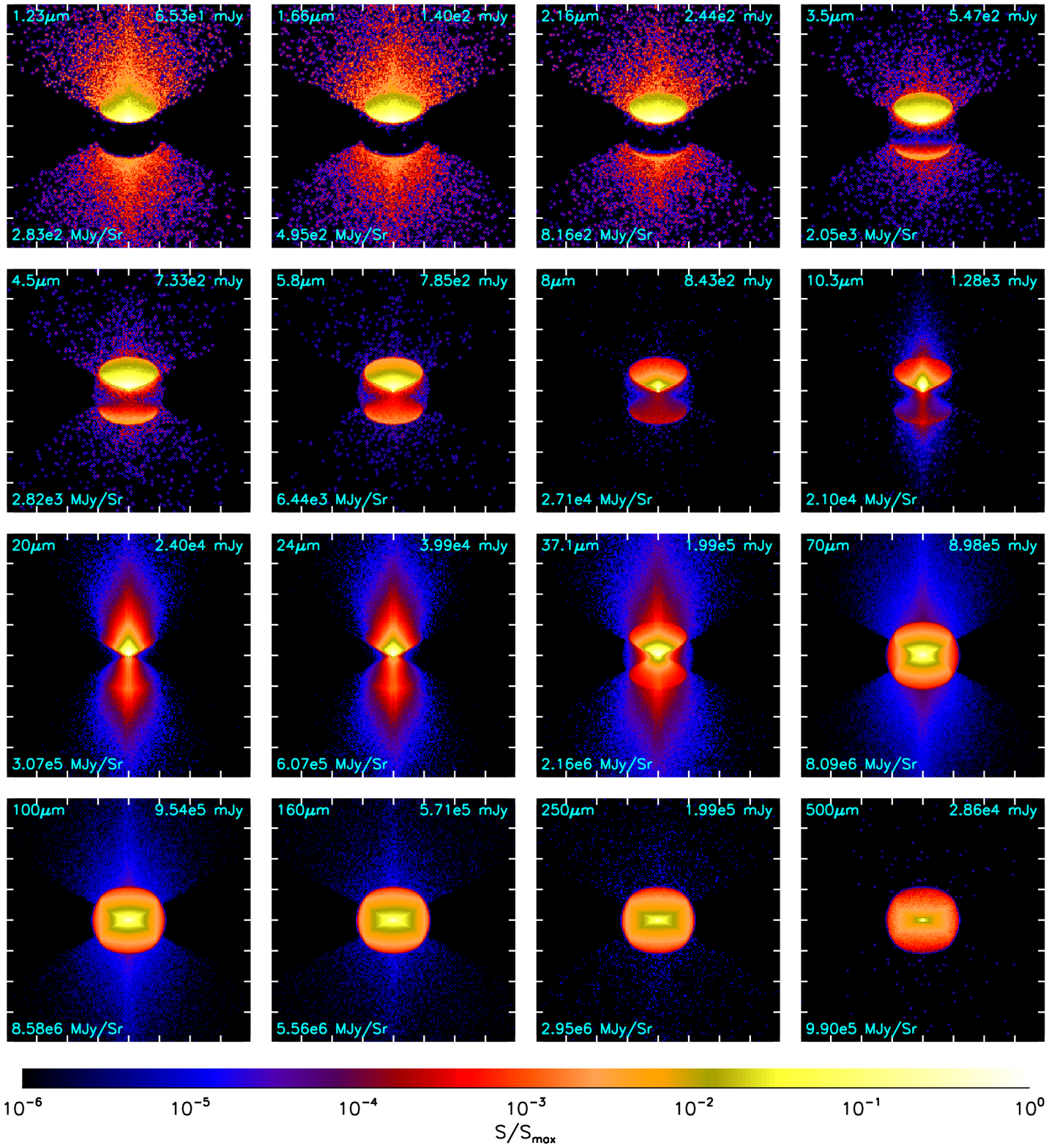}\\
\caption{Images for the fiducial model at an inclination of $60^\circ$, assuming a distance of 1 kpc, at different wavelengths. Each image has a field of view of 
$80\arcsec \times 80\arcsec$ ($80\arcsec$ corresponds to $8\times 10^4$ AU=0.4 pc), 
and is normalized to its maximum surface brightness, which is labeled in the bottom-left corner. The total fluxes are labeled in the top-right corners.}
\label{fig:img_60}
\end{center}
\end{figure*}

\begin{figure*}
\begin{center}
\includegraphics[width=0.97\textwidth]{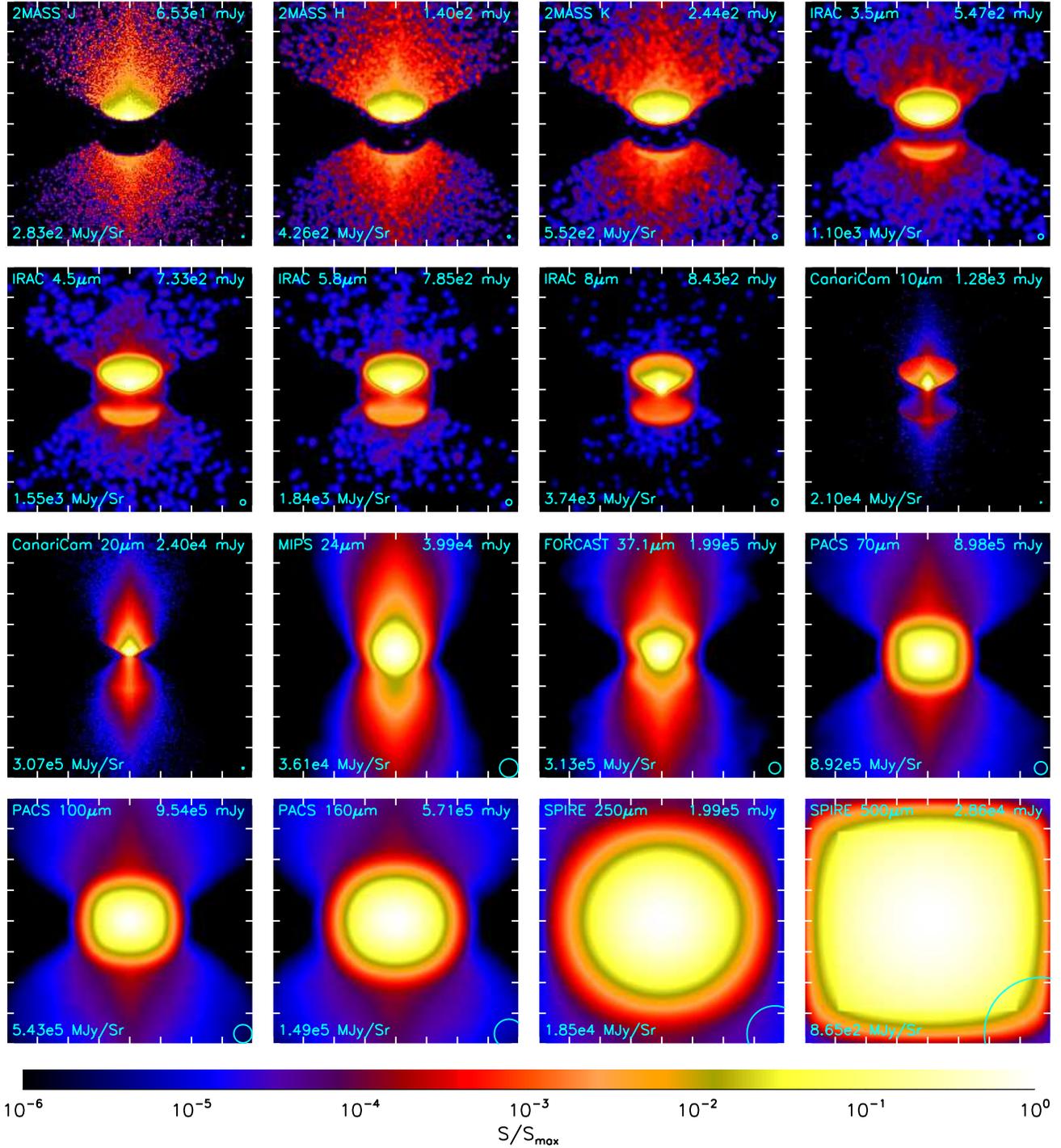}\\
\caption{Same as Figure \ref{fig:img_60}, except convolved with the beam of each instrument, labeled in the top-right and shown as red circles of radius equal to the HWHM on the bottom-right. The maximum surface brightness, which is dependent on the size of the beam, is labeled in the top-right corners.}
\label{fig:img_60_conv}
\end{center}
\end{figure*}

\begin{figure*}
\begin{center}
\includegraphics[width=0.97\textwidth]{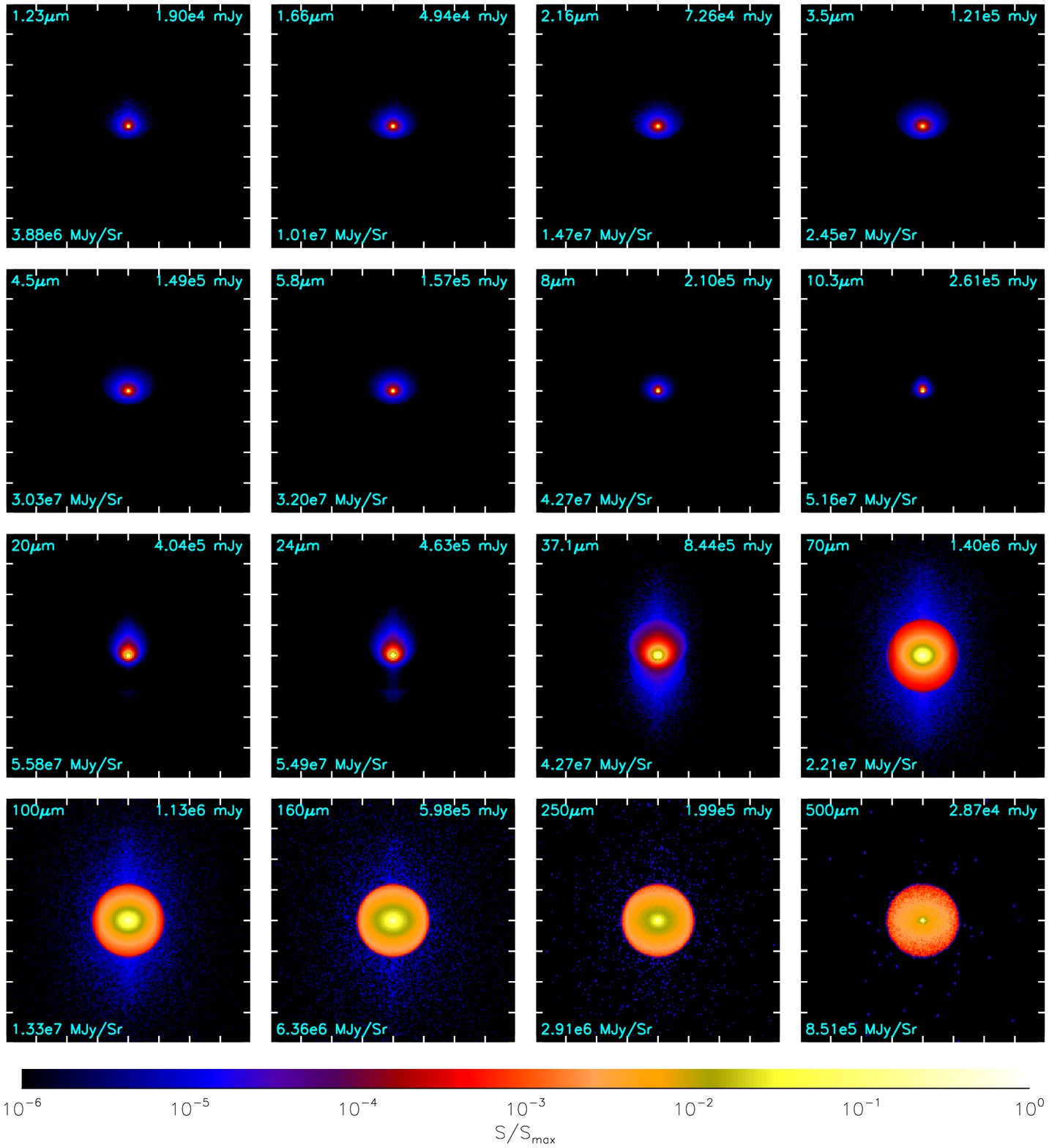}\\
\caption{Same as Figure \ref{fig:img_60}, except at an inclination of $30^\circ$.}
\label{fig:img_30}
\end{center}
\end{figure*}

\begin{figure*}
\begin{center}
\includegraphics[width=0.97\textwidth]{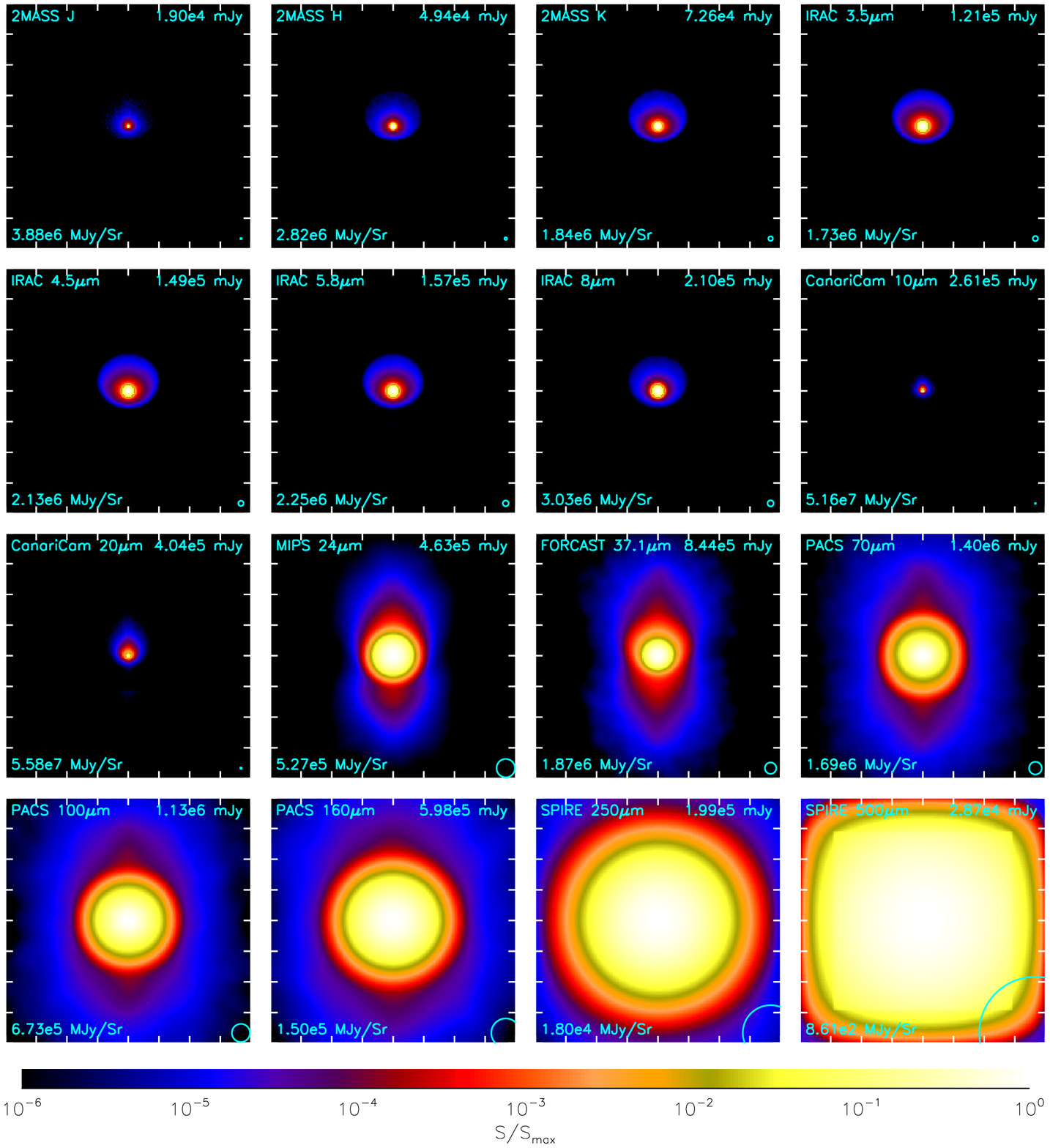}\\
\caption{Same as Figure \ref{fig:img_60_conv}, except at an inclination of $30^\circ$.}
\label{fig:img_30_conv}
\end{center}
\end{figure*}

\begin{figure*}
\begin{center}
\includegraphics[width=0.65\textwidth]{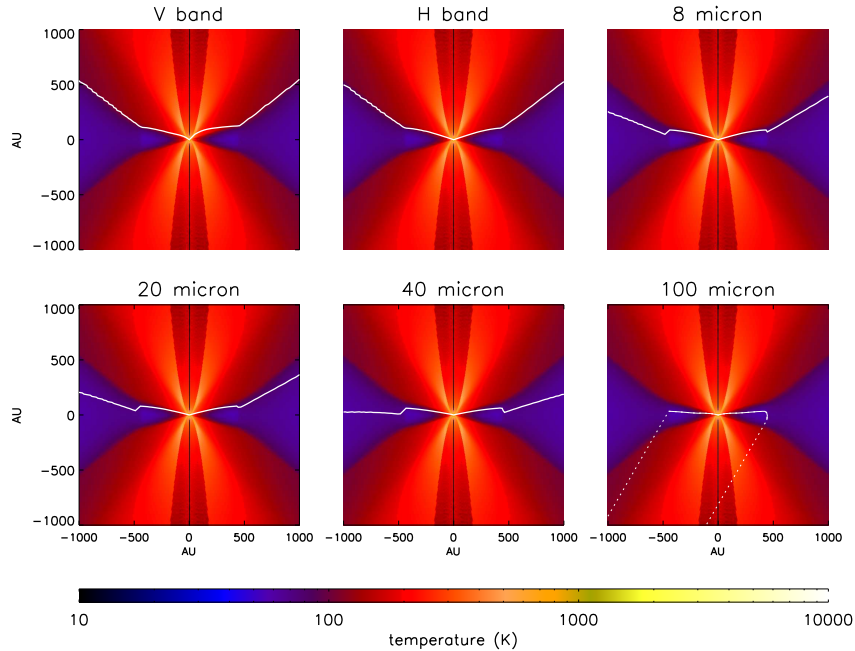}\\
\caption{The solid contours show positions of the 
$\tau=1$ surfaces
for different wavelengths seen from outside. In this case the observer is looking from the upper right at an angle of $30^\circ$ with respect to the axis.}
\label{fig:photosphere}
\end{center}
\end{figure*}

\begin{figure}
\begin{center}
\includegraphics[width=\columnwidth]{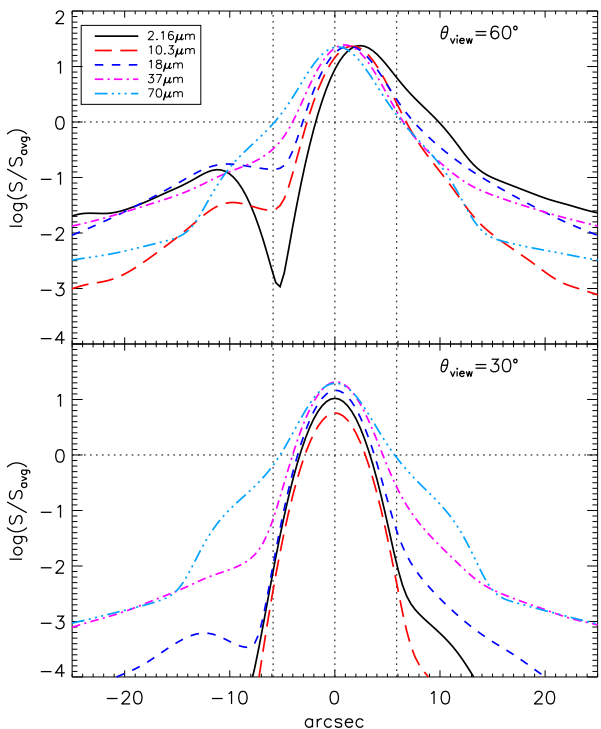}\\
\caption{Intensity distribution along the projected outflow axis. 
The surface brightness are convolved with a beam of FMHW of 4'' and normalized by the mean value of a 
$2\arcsec$ strip across the core along the axis at that wavelength. 
Different curves are for different wavelengths. The upper panel is at inclination of $60^\circ$ and the lower panel is at $30^\circ$. The vertical dashed lines on the left and right mark the positions of an offset of half the core radius to the center.}
\label{fig:fluxdist}
\end{center}
\end{figure}

\begin{figure*}
\begin{center}
\includegraphics[width=0.9\textwidth]{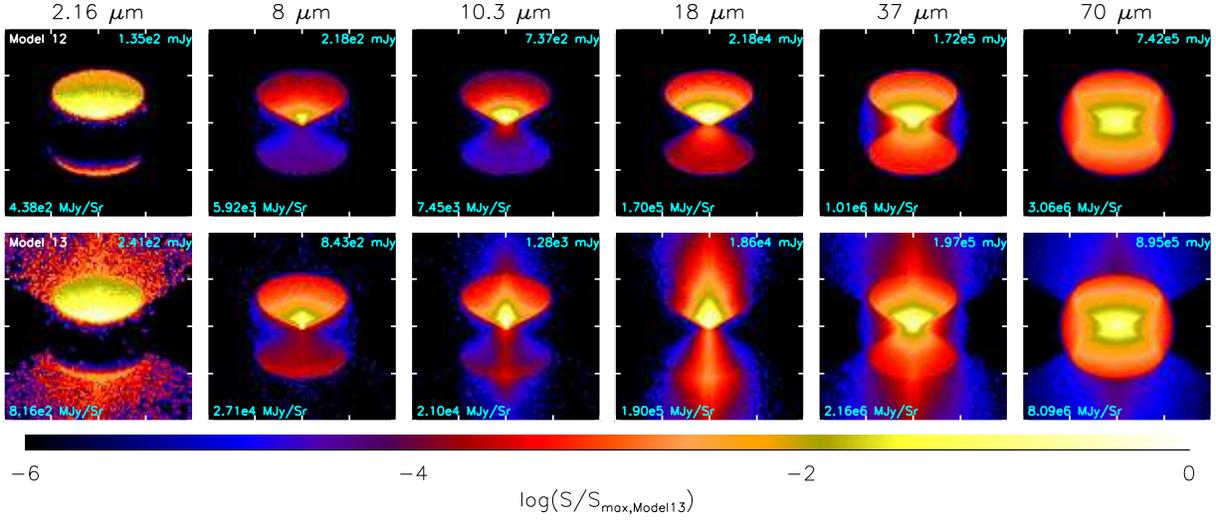}\\
\caption{Images at an inclination of $60^\circ$ and a distance of 1 kpc, comparing Model 12 and Model 13. The size of the images is 
$40\arcsec \times 40\arcsec$. 
At each wavelength, the images are normalized to the maximum surface brightness of Model 13. The total flux and the maximum surface brightness of each image are also labeled in the upper-right and lower-left corners.}
\label{fig:img_series}
\end{center}
\end{figure*}

\begin{figure*}
\begin{center}
\includegraphics[width=0.7\textwidth]{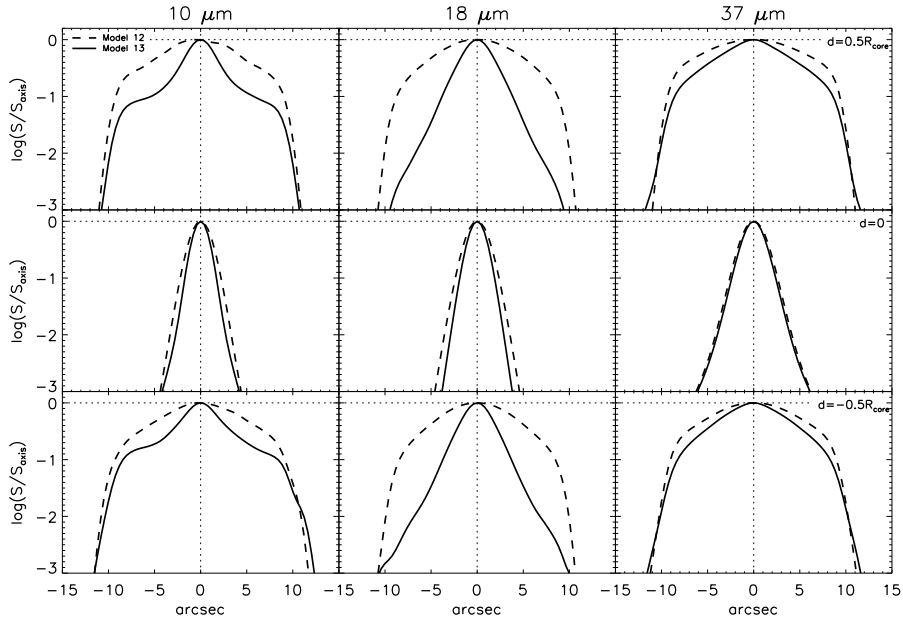}\\
\caption{Flux distribution along a strip crossing the core and perpendicular to the axis, comparing Model 12 and Model 13 at an inclination of $60^\circ$.
Fluxes are convolved with a beam of FMHW of 
$4\arcsec$
and normalized to the values on the axis. From top to bottom, the intensity profiles are along strips crossing the core with offsets to the center of $R_\mathrm{c}/2$ (the facing outflow), 0, and $-R_\mathrm{c}/2$ (the receding outflow). From left to right, different wavelengths are shown.}
\label{fig:fluxdisth}
\end{center}
\end{figure*}

\begin{figure*}
\begin{center}
\includegraphics[width=0.9\textwidth]{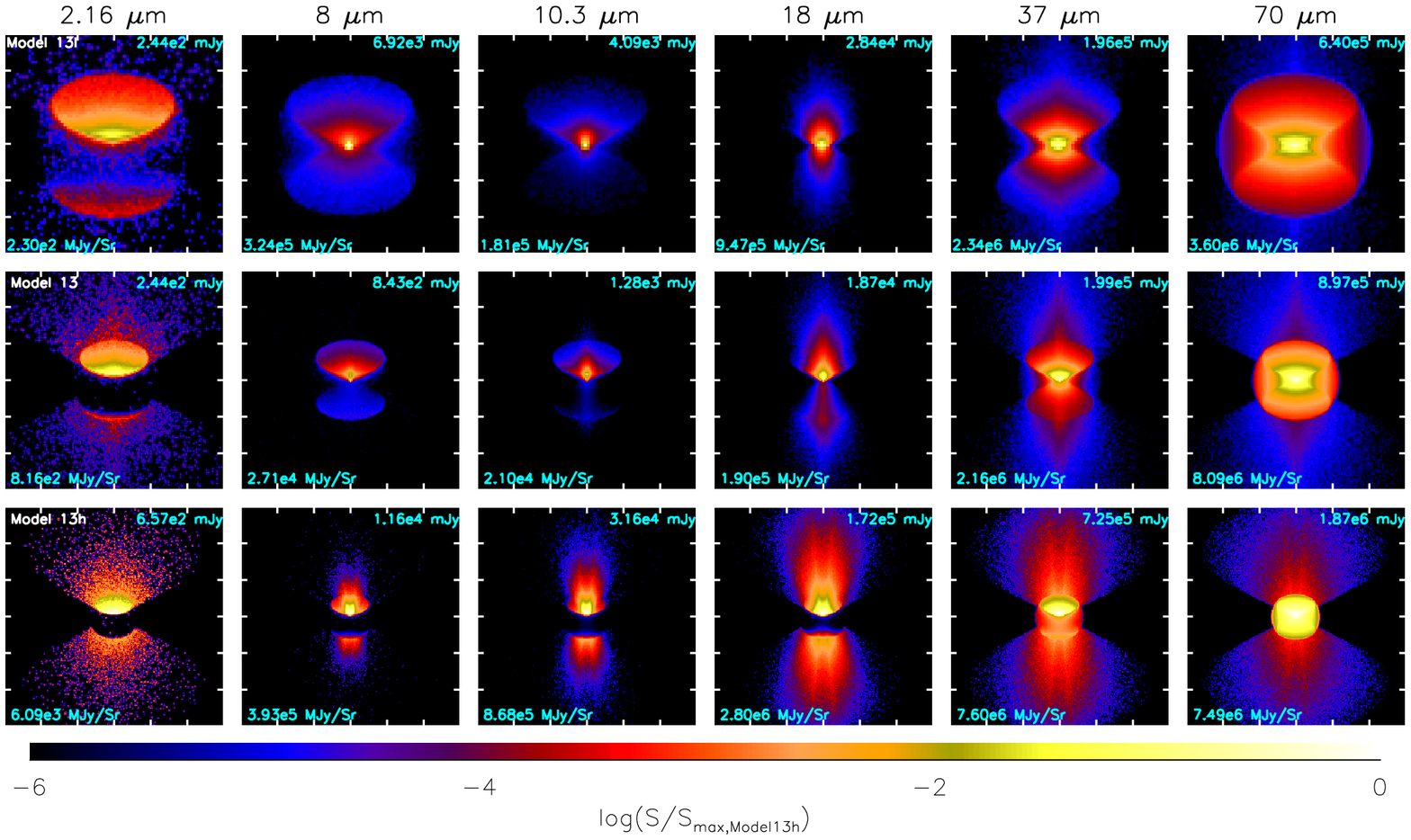}\\
\caption{Images of Model 13l ($\Sigma_\mathrm{cl}=0.316$ g cm$^{-2}$), Model 13 ($\Sigma_\mathrm{cl}=1$ g cm$^{-2}$), and Model 13h ($\Sigma_\mathrm{cl}=3.16$ g cm$^{-2}$) at an inclination of 60$^\circ$, assuming a distance of 1 kpc. The size of the images is 
$60\arcsec \times 60\arcsec$. 
Each image is scaled to the maximum surface brightness of Model 13h at that band with their own maximum brightness labeled in the bottom left corners. The total fluxes are labeled in the top right corners.}
\label{fig:img_sigma}
\end{center}
\end{figure*}

Figures \ref{fig:img_60}-\ref{fig:img_30_conv} present the images for
the fiducial model at inclinations of $60^\circ$ and $30^\circ$. We
show both resolved images and those after convolving with the PSFs of
16 different instrument bands. A distance of 1 kpc is assumed and no
foreground extinction is applied. Here, $5\times10^8$ photons are used
to produce images, but one still can see the noise due to the Monte
Carlo method in the extended outflow, especially in the bands with low
fluxes.

These results confirm the conclusion of our previous study: The
outflow cavity has a significant influence on the images up to
$70\mu$m, especially in the mid-IR, where the cavity wall (the inner part
of the accretion flow that is heated by the protostar) dominates the
morphology of the image.  At around 20 $\mu$m, the thermal emission
from the dusty outflow itself dominates the morphology.  At an
inclination of $60^\circ$, the receding side of the outflow is much
dimmer (by 1 to 2 orders of magnitude at 10 - 20 $\mu$m) than the
facing side. The source becomes more symmetric at longer wavelengths.
At $\sim 40 \mu$m, the receding outflow and cavity appear with a flux
within a factor of 10 of that from the approaching side.  The envelope
dominates the image at wavelengths $\gtrsim 70\mu$m.
In addition to the wide angle outflow, a much narrower jet-like
structure due to the denser region in the outflow close to the axis is
also seen at an inclination of $60^\circ$. There is also a trend that
from near-IR to $\sim 20 \mu$m, the outflow appears to be narrower.
Recall that we have assumed that the outflow extends to a radius of
$5R_\mathrm{c}$. This extended outflow material is brightest in the
near-IR because it is dominated by the light from the hot central
region that is scattered towards the observer. Note that our model
does not include the material of the self-gravitating clump that is
pressurizing the core.  A real source would most likely be embedded in
such a clump, and the appearance of the outflow outside the core would
most likely be affected by additional foreground extinction.

At an inclination of $30^\circ$, the star and disk could be seen
directly in the previous models (up to Model 11) because the outflow
cavity was treated as being optically thin. How are the locations of the
$\tau=1$ surfaces changed by the existence of the dust in the outflow
cavity? Figure \ref{fig:photosphere} shows the $\tau=1$ surfaces at
different wavelengths seen from outside of the source at an
inclination of $30^\circ$ for Model 13.  Even though there is dust in
the outflow, the observer may still see
down to the very base of the disk wind close to the protostar and disk
even in the optical band.  The $\tau=1$ surface basically lies on the
surface of the accretion disk in the near
infrared bands, and reaches deeper as
the wavelength increases.  
At inclinations large enough that the line of sight passes through the envelope,
the source is relatively unobscured at wavelengths $\ga 100 \mu$m.  
With the angular resolution achieved by large sub-mm interferometers such as
ALMA, such a disk (with a size of $\sim$ 1000~AU)
could be readily resolved at a distance of several kpc.
 
Figure \ref{fig:fluxdist} shows the intensity profiles along the
outflow axis at inclinations of $60^\circ$ and $30^\circ$. The profiles
are convolved with a beam with FHWM of 4 \arcsec\
and normalized by the mean intensity at that wavelength
of a $2\arcsec$ wide strip along the axis across the core. 
This strip extends to where the intensity drops to $10^{-6}$ of the maximum intensity.
As discussed above, at an inclination of $60^\circ$, as one goes from
shorter to longer wavelengths, the profile becomes more symmetric and
the brightest point moves closer to the protostar.  The extended
facing outflow ($\theta>0$) is brightest in the near-IR; the
normalized brightness at 70 $\mu$m is similar to that at 10 $\mu$m.
The opposite side of the outflow can be best seen in the far-IR. At an
inclination of $30^\circ$, the profiles are much more symmetric.  The
degree of symmetry in the brightness profile appears to be a promising
indicator of the inclination of observed massive protostars.

Figure \ref{fig:img_series} compares the images of Model 12 (dust-free
outflow) and Model 13 (dusty outflow), showing that the dust makes
both sides of the outflow cavities brighter at all the wavelengths we
considered 
(which are between 2 and 70~$\mu$m)
except at 18 $\mu$m, where the emission of the dusty
outflow dominates and the flux is more concentrated to the central
region. The dust in the disk wind also causes the outflow to appear narrower
at $10$ and 18 $\mu$m than in the near-IR.
Figure \ref{fig:fluxdisth} shows similar intensity profiles as Figure
\ref{fig:fluxdist}, but perpendicular to the axis
at an inclination of 60$^\circ$.
At 10 and 18 $\mu$m, the profiles of the bright component of the
emission are narrower in Model 13.  The difference between the two
models becomes small for $\lambda \ga 37 \mu$m.

The effects of the surface density of the surrounding clump,
$\Sigma_\mathrm{cl}$, are shown in Figure \ref{fig:img_sigma}. The
surface brightnesses are normalized to the maximum value of Model 13h
at each wavelength. Model 13h has an envelope with higher extinction,
which causes a higher contrast between the two sides of the outflow
than in Model 13 or Model 13l. The sides becomes more symmetric when
$\Sigma_\mathrm{cl}$ is lower (Model 13l). Also, we can see that the
outflow cavity is brighter at mid-IR wavelengths (8 and 10.3 $\mu$m)
in the two cases with higher extinction, as opposed to the model with
the lowest extinction, for which the emission is concentrated at the
center.

\section{Summary}
\label{sec:summary}

We have constructed radiation transfer models for individual massive
star formation, with a 60~$M_\odot$ initial core and an 8~$M_\odot$
protostar forming at the center.

1. Compared with Paper I, our model is improved in two important
aspects: the effects of the outflow, inflow and time dependence on the
disk surface-density profile are considered, and a detailed
distribution of dust and gas in the outflow cavity has been studied in
a self-consistent way. While most of the improvements in the disk
model (except for the reduction in the accretion energy, and hence the
luminosity, due to driving the wind) have only minor influence on the
results, the inclusion of dust in the outflow cavity can strongly
affect both the SEDs and the images.

2. We have presented a simple analytic model for the density
distribution in a disk wind that is based on the \citet{BP82}
solution, generalized to allow for an arbitrary outer boundary.

3. SEDs of the series of models at inclinations of 30$^\circ$ and
60$^\circ$ are presented. Including dust in the outflow cavity
significantly influences the SEDs at both inclinations. The dust in
the outflow suppresses emission at short wavelengths, while
contributing a significant amount of near- and mid-IR emission. The
disk also becomes warmer due to the re-radiated light from the outflow
material, producing higher mid- and far-IR emission.  Close to a
face-on view, at inclinations from $\sim 10^\circ$ to $\sim 30^\circ$,
the SED from near-IR to about 60 $\mu$m is relatively flat, which may
be used to identify face-on MYSOs. The 10 $\mu$m silicate feature
appears only if the mid-IR flux from the disk is high enough (e.g. due
to lower extinction of the envelope) so that it is not buried in the
flux from the envelope, which indicates that the depth of this feature
is determined both by the total optical depth at 10$\mu$m and by the
relative brightness of the disk and the envelope.

4. We have presented images of the fiducial model for different
observational bands, both resolved and convolved with the resolution
beams of the instruments. As discussed in Paper I, the outflow cavity
is the most significant feature on the images up to 70 $\mu$m. The
outflow cavity facing the observer may dominate the mid-IR flux and
determine the morphology, while the opposite side can be seen only at
wavelengths $\gtrsim$ 30 $\mu$m. Even though the outflow cavity is
wide, at 10 to 20 $\mu$m, the dust distribution in the disk wind can
make the outflow appear narrower in the observed images. The
distribution of the dust in the outflow affects the intensity profiles
both along the axis and perpendicular to the axis. Comparison of such
profiles with observation may potentially determine the properties of
the outflow. At a lower inclination (more face-on), the disk becomes
visible even in the optical bands, although foreground extinction from
the natal clump and the larger-scale intervening ISM may limit
detection of such emission in practice. 

5. We studied the effects of the surface density ($\Sigma_\mathrm{cl}$
and hence the ambient pressure, of the environment in which the core
is embedded. Higher $\Sigma_\mathrm{cl}$ makes the core denser,
leading to a higher accretion rate. It affects the SED in two ways:
the higher accretion rate leads to higher luminosity, and the denser
envelope (and outflow) leads to higher extinction, which can suppress
emission at mid-IR and shorter wavelengths. The higher extinction also
affects the images, making higher contrast between the two sides of
the outflow, while in case of lower extinction, the two sides are more
symmetric and the central region is more prominent.

6. We studied how the inferred luminosity depends on the viewing angle
(the flashlight effect)
The luminosity inferred from the flux from a nearly face-on view can
be higher than that from an edge-on view by 2 orders of magnitude in
the near-IR and mid-IR. The inferred luminosity from a nearly face-on
view is typically higher than the true IR luminosity by a factor of
about 2.  In the mid- and far-IR up to $\sim$ 70 $\mu$m, the face-on
fluxes, $\nu F_\nu$, can also give a good prediction of the bolometric
luminosity.  When integrated from a SED, the inferred bolometric
luminosity is higher than the true value by a factor of 3 if looking
down the outflow cavity, i.e. most of the radiation escapes from the
polar direction.

\acknowledgements We thank Barbara Whitney for helpful discussions
and for providing the latest version of her radiation transfer code.
YZ acknowledges support from a Graduate School
Fellowship from the Univ. of Florida.  JCT acknowledges support from
NSF CAREER grant AST-0645412; NASA Astrophysics Theory and Fundamental
Physics grant ATP09-0094; and NASA Astrophysics Data Analysis Program
ADAP10-0110.  The research of CFM is supported in part by the NSF
through grants AST-0908553 and AST-1211729,
and NASA through AFTP grant NNX09AK31G.

\appendix

\section{A. An Improved $\alpha$-Disk Solution}
\label{app:disk}

Here we summarize the formulae we use to include the effects of
outflow/inflow and of a massive, growing disk in an Shakura-Sunyaev
$\alpha$-disk model. Following the approach in standard textbooks
(e.g. \citealt[]{FKR02}), we write the conservation of the mass and
angular momentum for an annulus at radius of $\varpi$ as
\begin{eqnarray}
&&\varpi\ppbyp{\Sigma}{t}+\frac{\partial}{\partial \varpi}(\varpi\Sigma v_\varpi) 
+ \varpi \ppbyp{\Sigma_w}{t} -  \varpi \ppbyp{\Sigma_\mathrm{in}}{t} = 0,\\ 
&&\varpi\frac{\partial}{\partial t}(\Sigma \varpi^2 \Omega)  
+\frac{\partial}{\partial \varpi}(\varpi\Sigma v_\varpi \varpi^2 \Omega)-
\frac{1}{2\pi}\frac{\partial G_\mathrm{T}}{\partial \varpi}+\varpi j_w\ppbyp{\Sigma_w}{t} 
-\varpi j_\mathrm{in}\ppbyp{\Sigma_\mathrm{in}}{t}  = 0, 
\end{eqnarray}
where $\Sigma$ and $\partial{\Sigma}/\partial t$ are the surface density of the disk
at $\varpi$ and its rate of growth, $\partial{\Sigma}_w/\partial t$ and
$\partial{\Sigma}_\mathrm{in}/\partial t$ 
are the surface mass loss/loading rates into
the wind/onto the disk at $\varpi$, 
$v_\varpi$ is the radial velocity of the
accretion flow, $G_\mathrm{T} = 2\pi \varpi \nu \Sigma \varpi^2 (d\Omega/d\varpi)$ is the
viscous torque, $j_w$ and $j_\mathrm{in}$ are the specific angular
momenta of the wind and the inflow, and $\Omega$ is the angular
frequency. Integration from $r_*$ gives
\begin{eqnarray}
&&\mda(\varpi)\equiv -2\pi \varpi\Sigma v_\varpi=
\pbyp{t}\left[{m}_d(\varpi)
+{m}_w(\varpi)-{m}_\mathrm{in}(\varpi)\right]
+\dot{m}_*,\label{eq:mass}\\
&&(-2\pi \varpi) (\nu \Sigma) \left(\varpi^2 \frac{\partial \Omega}{\partial \varpi}\right)
= (\varpi^2\Omega)\mda - \dot{m}_* (r_*^2\Omega_*) 
-\pbyp{t}\left[ {J}_d(\varpi)+{J}_w(\varpi)-{J}_\mathrm{in}(\varpi)\right],
\label{eq:ang}
\end{eqnarray}
where subscript * indicates values on the surface of the star (i.e. the inner boundary of the disk), and
\begin{eqnarray}
&&\ppbyp{m_d(\varpi)}{t}=\int^\varpi_{r_*}2\pi \varpi' \ppbyp{\Sigma}{t}\, d\varpi',
\qquad \ppbyp{m_w(\varpi)}{t}=\int^\varpi_{\varpi_*}2\pi \varpi' \ppbyp{\Sigma_w}{t}\, d\varpi',\\
&&\ppbyp{m_\mathrm{in}(\varpi)}{t}=\int^\varpi_{r_*}2\pi \varpi' \ppbyp{\Sigma_\mathrm{in}}{t}\, d\varpi',\\
&&\ppbyp{J_w(\varpi)}{t}=\int_{r_*}^\varpi 2\pi \varpi' j_w\ppbyp{\Sigma_w}{t}  d\varpi',
\qquad \ppbyp{J_\mathrm{in}(\varpi)}{t}=\int_{r_*}^\varpi 2\pi \varpi' 
j_\mathrm{in} \ppbyp{\Sigma_\mathrm{in}}{t}  d\varpi',\\
&&\ppbyp{J_d(\varpi)}{t}=\int_{r_*}^\varpi 2\pi \varpi' \frac{\partial}{\partial t} 
(\Sigma \varpi'^2 \Omega) d\varpi'
=\int_{r_*}^\varpi 2\pi \varpi' \ppbyp{\Sigma}{t} (\varpi'^2 \Omega) d\varpi'
+\int_{r_*}^\varpi 2\pi \varpi' \Sigma \left(\varpi'^2 \ppbyp{\Omega}{t}\right) d\varpi'.
\end{eqnarray}

We can write Equation (\ref{eq:ang}) as
\begin{eqnarray}
\nu \Sigma & = & \left[ \mda(\varpi)-\dot{m}_* \left(\frac{m_*}{m_*+m_d(\varpi)}\right)
^\frac{1}{2} \left(\frac{r_*}{\varpi}\right)^\frac{1}{2} - 
\frac{\partial\left[{J}_d(\varpi)+
{J}_w(\varpi)-{J}_\mathrm{in}(\varpi)\right]/\partial t } {\sqrt{G (m_*+m_d(\varpi))\varpi}} \right]
\nonumber\\
& & \left/\left[3\pi-\frac{2\pi^2 \Sigma \varpi^2}{m_*+m_d(\varpi)}\right] \right. ,\label{eq:nusigma}
\end{eqnarray}
compared with a disk with constant accretion rate through $R$:
\begin{equation}
\nu \Sigma =\frac{\dot{m}_*}{3\pi}\left[1-\left(\frac{r_*}{\varpi}\right)^{1/2}\right]\label{eq:nusigma_old}.
\end{equation}

The equation set for a Shakura-Sunyaev disk is:
\begin{equation}
\left.
\begin{array}{l}
1.\, \rho=\Sigma/H \\
2.\, H=c_s/\Omega \\
3.\, c_s^2=P/\rho \\
4.\, P=\frac{\rho k}{\mu m_\mathrm{p}}T_c 
+\frac{4\sigma}{3c}T_c^{4}
\\
5.\, \frac{4\sigma}{3\tau} T_c^4=\frac{1}{2}
\nu\Sigma
(\varpi\Omega')^2 \\
6.\, \tau=\Sigma \kappa(\rho,T_c) \\
7.\, \nu=\alpha c_s H \\
\end{array}
\right\}\label{eqs}
\end{equation}
The radiation pressure term is small for the current fiducial model, and
thus neglected here.

With the boundary conditions $m_d=f_d m_*$, $\dot{m}_d(\varpi_d)=f_d \dot{m}_*$,
$\dot{m}_w(\varpi_d)=f_w \dot{m}_*$, 
$\dot{m}_\mathrm{in}(\varpi_d)=(1+f_d+f_w) \dot{m}_*$,
and $\partial{\Sigma}_w/\partial t$, $\partial{\Sigma}_\mathrm{in}/\partial t$, 
$j_w$ and
$j_\mathrm{in}$ given by the solutions of the wind and inflow, at any moment $t$, 
once we have a profile of 
$\partial{\Sigma}(\varpi)/\partial t$, 
we can calculate $\Sigma(\varpi)$ and other disk profiles ($T_c$, $H$, etc.)
by combining Eq. (\ref{eq:mass}), Eq. (\ref{eq:nusigma}) and Eqs. (\ref{eqs}).  
In fact, we set 
$\partial{\Sigma}/\partial t=0$ at first, and calculate $\Sigma(\varpi)$ at two close
moments in time ($t$ and $t+dt$), and use these two $\Sigma$ profiles to
re-estimate 
$\partial{\Sigma}(\varpi)/\partial t$, and iterate until the two converge.
Note the size of the disk $\varpi_d$ can also vary at these two moments,
i.e. the effect of an increasing disk size is also consistently included.
We also use the surface density of the inner adjacent annulus to estimate
$\Sigma$ in the r.h.s of Equation (\ref{eq:nusigma}) to make the
calculation easier.

The effect of the increasing size of the disk on the disk surface density can
be estimated as below. The total derivative of the disk mass can be written as
\begin{eqnarray}
\dot{m}_d(\varpi_d) & = & \int_{r_*}^{\varpi_d}2\pi\frac{\partial \Sigma(\varpi)}{\partial t} \varpi d\varpi
+2\pi\Sigma(\varpi_d) \varpi_d \dot{\varpi}_d\nonumber\\
& = & \frac{\partial m_d(\varpi)}{\partial t}\Big|_{\varpi=\varpi_d}+2\pi\Sigma(\varpi_d) \varpi_d \dot{\varpi}_d.
\end{eqnarray}
The size of the disk $\varpi_d\propto m_d(\varpi_d)^{1/(3-k_\rho)}$
(Eq. 13 in Paper I), so that
$\dot{\varpi_d}/\varpi_d=1/(3-k_\rho)[\dot{m}_d(\varpi_d)/m_d(\varpi_d)]$.
Then we can write
\beq
\frac{\partial m_d(\varpi)}{\partial t}\Big|_{\varpi=\varpi_d}
=\dot{m}_d(\varpi_d)\left[1-\frac{2}{3-k_\rho}\frac{\Sigma(\varpi_d)}{\bar{\Sigma}}\right],
\eeq
where $\bar{\Sigma}=m_d(\varpi_d)/(\pi\varpi_d^2)$. We can see that because the size of
the disk is increasing, the mass of the disk inside $\varpi_d$ may either increase or decrease
depending on the ratio of the surface density at the boundary of the disk to the average surface
density of the disk. In our fiducial case, $\Sigma(\varpi_d)/\bar{\Sigma}\gtrsim 1$, then $\partial m_d(\varpi)/\partial t\big|_{\varpi=\varpi_d}<0$, i.e. the disk mass inside $\varpi_d$ is decreasing.

\section{B. General Formulation of the Density Distribution of the Outflow}
\label{app:outflow}

We wish to develop an approximate model for a disk wind that will fill the
cavity described in Paper I. Protostellar disk winds are believed to be
hydromagnetic in origin (K\"onigl \& Pudritz 2000),
but we shall not explicitly consider the magnetic field here.
Our goal is to determine the density distribution in the wind, and
we shall make use of the results for the disk wind evaluated by \citet[BP wind]{BP82} 
and refined by \citet[]{CL94}.
Our results asymptotically approach those of \citet[]{Ostriker97} for self-similar cylindrical winds.
We assume that the disk is thin and Keplerian, that the wind is steady
and that it is launched 
from the surface of the disk
in a self-similar fashion.

We define the base of the wind as being at the point at which 
the vertical velocity $v_z$ reaches some fixed, small
fraction of the local Keplerian velocity
$v_K$; we label this velocity $\vzo$. 
The precise value of the ratio $\vzo/v_K$
is irrelevant, but the ratio must be sufficiently small that
the base of the wind is 
close to the surface of the disk. 
We allow for the finite thickness of the disk, but we assume that it is small so
that the velocity normal to the surface of the disk is approximately in the $z$ direction.
Let $\vp\equiv r\sin\theta$
be the cylindrical radius of a streamline and $\vpo$ denote the value of
$\vp$ at the base of the wind. 
The corresponding values of the innermost and outermost radii
at the base of the wind are $\vpco$ and $\vpmo$, respectively.
The height of the base of the wind, i.e. the surface of the disk, is denoted
by $z_d(\vpo)$.
For a streamline starting from [$\vpo$, $z_d(\vpo)$], we define the effective height
to be 
$z'(\vpo)=z-z_d(\vpo)$.
We assume that there is a central cavity in the wind created by the magnetic field
threading the star (\citealt[]{Shu95}); 
let $\vpc[z'(\vpco)]$ be the radius of this cavity. 
We further assume that the wind is confined by the ambient medium,
which sets $\vpm[z'(\vpmo)]$, the radius of the outer boundary of
the wind; we shall specifiy this below.
We introduce the normalized radius at the base of the wind,
\beq
x_0\equiv \frac{\vpo}{\vpco}.
\eeq
We then have $v_K=v_{Kc}x_0^{-1/2}$ and $\vzo=\vzco x_0^{-1/2}$,
where $v_{Kc}=(Gm_*/\vpco)^{1/2}$ and $\vzco$ are measured on the innermost streamline.
Let $\rho_0(\vpo)$ be the density at the base of the wind
for the streamline originating at $\vpo$.
We assume that disk wind is launched in a self-similar manner, so that
$\rho_0(\vpo)$ varies as a power-law
in radius,
\beq
\rho_0(\vpo)=\rho_{c0} \left(\frac{\vpo}{\vpco}\right)^{-q}
\equiv \rho_{c0}x_0^{-q}\; ,
\eeq
where $\rho_{c0}$ is the density at the base of the wind along
the innermost streamline.

The mass-loss rate from a ring of width $d\vpco$ on both sides of the surface of the disk is
\beqa
d\mdotw&=&4\pi\vpo\rho_0\vzo d\vpo,\\
&=& 4\pi\vpco^2\rho_{c0}\vzco x_0^{\frac 12 -q} dx_0,
\label{eq:dmdwo}
\eeqa
where we have made use of assumed power-law scalings of the density and velocity.
The total mass-loss rate from the disk is then
\beq
\mdotw=4\pi\vpco^2\rho_{c0}\vzco I_w(\xmo)\ln\xmo,
\label{eq:mdw}
\eeq
where
\beq
I_w\equiv\frac{1}{\ln\xmo}\left(\frac{\xmo^{\frac 32-q}-1}{\frac 32 -q}\right).
\label{eq:iw}
\eeq
BP winds have $q=\frac 32$, and in this case $I_w=1$.

The mass loss rate through rings of width $d\vp$ at a height $z'$ above and below the 
surface of 
the disk is 
approximately
the same as that in the plane of the disk,
\beq
d\mdotw=4\pi\vp\rho v_z d\vp.
\eeq
Let
\beq
V(\vpo,z')\equiv\frac{v_z(\vpo,z')}{v_K(\vpo)}\label{eq:vv}
\eeq
be the ratio of the vertical velocity to the Keplerian velocity at the footpoint,
and let
\beq
\eta\equiv\left.\frac{\partial\ln\vpo}{\partial\ln\vp}\right|_{z'}
\label{eq:eta}
\eeq
measure the change in the width of the ring.
Equating the mass loss rate at a height $z'$ to that in Equation (\ref{eq:dmdwo}) and using Equation
(\ref{eq:mdw}) to eliminate $\rho_{c0}$, we then find that the density in the wind is
\beq
\rho=\left(\frac{\mdotw}{4\pi \vkc I_w\ln\xmo}\right)
\frac{\eta x_0^{2-q}}{V \varpi^2}.
\label{eq:rho1}
\eeq
This is a generalization of the results of \citet{Ostriker97} for cylindrical
winds, 
which have $\eta=1$. 
Note that if the variation of $x_0$ is
small 
(compared to that of 
$\varpi$),
then Equation (\ref{eq:rho1}) implies
that $\rho\propto \vp^{-2}$ (\citealt[]{MM99}). In
particular, this is the case for 
X-winds (\citealt[]{Shu94,Shu95}), for which $x_0\simeq 1$.

\subsection{B1. Approximate Streamlines}

In order to complete our estimate of the density in the wind, we need to specify the streamlines.
We describe the streamlines in terms of the dimensionless radius,
\beq
R(\vpo, z')\equiv\frac{\vp(\vpo,z')}{\vpo},
\eeq
and the dimensionless height,
\beq
Z(\vpo,z')\equiv\frac{z'}{\vpo}.
\eeq
The innermost and outermost streamlines are $R_c(Z)=\vpc(Z)/\vpco$ and
$\rmax(Z)=\vpm(Z)/\vpmo$, respectively. 
We assume that the streamlines emanating from points in the
disk are a weighted mean of the innermost and outermost streamlines:
\beq
\varpi(\vpo,Z)=\vpc(Z)^{1-\delta}\vpm(Z)^\delta.
\label{eq:vpinterp}
\eeq
Since this relation applies at the surface of the disk ($Z=0$), one finds that
\beq
\delta=\frac{\ln x_0}{\ln\xmo}.
\eeq
Thus, for the innermost streamline ($x_0=1$), Equation (\ref{eq:vpinterp}) gives $\vp=\vpc$,
and for the outermost one ($x_0=\xmo$) it gives $\vp=\vpm$, as it should. The {\it ansatz} we
have adopted thus provides a smooth transition from the innermost to the outermost streamline as the footpoint
moves from $\vpco$ to $\vpmo$. In terms of the dimensionless radius, we have
\beq
R(Z)=R_c(Z)^{1-\delta}\rmax(Z)^\delta.
\label{eq:rapprox}
\eeq
It is important to note that in both this expression and in Equation
(\ref{eq:vpinterp}), the radius of the innermost and outermost streamlines are evaluated
at the same value of the normalized height, $Z$, not at the same height above the disk, $z'$.

In general, define the characteristic wind density as
\beqa
\rhoch&=&\frac{\mdotw}{4\pi\vpco^2 \vkc I_w\ln\xmo},\\
&=&\frac{4.3\times 10^{-11}}{I_w\ln\xmo}\left(\frac{10^{12}\mbox{ cm}}{R_*} \right)^{3/2} 
\left(\frac{M_\odot}{m_*} \right)^{1/2}
\left(\frac{\mdotw}{10^{-4}\;M_\odot\;\mbox{yr\e}}\right)~~~~\mbox{g cm\eee}.
\eeqa
Along a given streamline, the wind density (Eq. \ref{eq:rho1}) is then
\beq
\rho(\vpo,z')=\rhoch\left[\frac{\eta}{V(Z) R^2(x_0,z') x_0^{q}}\right].
\label{eq:rho2}
\eeq
The quantity $\eta$, which measures the rate at which the streamlines separate, is
given by Equation (\ref{eq:eta}), which can be expressed as
\beqa
\eta^{-1}&=&-(1-\delta)\ppbyp{\ln\vpc}{\ln Z}-\delta \ppbyp{\vpm}{\ln Z}+\ln\left(\frac{\vpm}{\vpc}\right)\;
\ppbyp{\delta}{\ln\vpo},\\
&=&-(1-\delta)\frac{d\ln R_c(Z)}{d\ln Z}-\delta\,\frac{d\ln \rmax(Z)}{d\ln Z} +1 +\frac{\ln[\rmax(Z)/R_c(Z)]}{\ln\xmo},
\eeqa
where all the partial derivatives are evaluated at constant $z'$. Note that if the streamlines are self-similar,
so that $R$ depends only on $Z$, and as a result $R_c(Z)=\rmax(Z)$, this reduces to
\beq
\eta^{-1}=1-\frac{d\ln R}{d\ln Z}~~~~~\mbox{(self-similar winds)}
\eeq
(Contopoulos \& Lovelace 1994).

\subsection{B2. BP-like Winds}

We now focus on winds that are similar to those considered by BP. 
Such winds emanate from a disk with a density distribution
that varies as $\vpo^{-q}$ with $q=\frac 32$, and as a result $I_w=1$ (Eq. \ref{eq:iw}).
For the particular numerical
example BP considered, the dimensionless radius is given by\footnote{
An expression that is more accurate at small $Z$ is $R\simeq 1+4.67\ln(1+0.36 Z +3.4\times 10^{-4}Z^3);$
in particular, this expression gives the correct value of $dR/dZ$ at $Z=0$, where the wind is launched.
The physical conditions in the wind-launching region are likely to be much more complicated
than given by the BP solution, however, so this additional accuracy is only apparent.}
\beq
R\simeq 1+14\ln(1+0.07Z),
\label{eq:rbp}
\eeq
which agrees with the numerical values given by \citet{CL94} to within
about 10\% for $Z\leq 1246$, beyond which the wind begins to contract toward the axis.
For larger values of $Z$, this approximation indicates that the flow is highly collimated, which is consistent
with observations of jets from low-mass protostars (\citealt{MO07}).
The dimensionless velocity (\ref{eq:vv}) in this wind is given by
\beq
V(Z)\simeq \ln(1.01+5Z^{0.8}),
\label{eq:v}
\eeq
which agrees with \citeapos{CL94}
numerical values to within about 10\%, except for very small $Z$;
we have inserted 1.01 instead of 1 in this expression in order to avoid the singularity
that would otherwise occur in the density at $Z=0$. If both the innermost and outermost
streamlines are BP streamlines, then Equation (\ref{eq:rho2}) gives an approximation
for the density in a BP wind.

In reality it is likely that the outermost streamline is determined by boundary conditions,
and we can use the analysis above to estimate the density in the resulting non-self-similar wind.
As in Paper I, we assume that the outermost streamline of the wind
corresponds to the innermost streamline of the accretion flow.
\citet{Ulrich76} developed a simple model for a rotating infall characterized
by a disk radius $\varpi_d$. Gas originating far from the protostar at a polar angle
$\theta_0$ follows a trajectory in polar coordinates $(r,\theta)$ given by
\beq
\frac{r}{\varpi_d}=\frac{\mu_0(1-\mu_0^2)}{\mu_0-\mu},
\eeq
where $\mu=\cos\theta$ and $\mu_0=\cos\theta_0$. 
Let $\twe$ be the largest angle at which the wind can escape,
evaluated far from the disk. 
If this
outermost streamline of the wind starts at ($\vpmo$, $z_{d,\mathrm{max}}$) 
and corresponds to a streamline of the accretion flow,
then $\theta_0=\twe$,
and $z_{d,\mathrm{max}}$ can be determined by
\begin{equation}
\vpmo=\left[\varpi_d^2(1-\mu_0^2)^2+z_{d,\mathrm{max}}^2(1-\mu_0^2)/\mu_0^2
+2\varpi_d z_{d,\mathrm{max}}(1-\mu_0^2)/\mu_0 \right]^{1/2}.\label{eq:zdmax}
\end{equation}
One can then show that
\beq
\rmax=\left[Z_{\rm max}^2\left(\frac{1-\mu_0^2}{\mu_0^2}\right)+
2Z_{\rm max}\left(\frac{\varpi_d}{\vpmo}\right)\left(\frac{1-\mu_0^2}{\mu_0}\right)+
2Z_{\rm max}\left(\frac{z_{d,\mathrm{max}}}{\vpmo}\right)\left(\frac{1-\mu_0^2}{\mu_0^2}\right)+
1 \right]^{1/2},\label{eq:Rmax}
\eeq
where $Z_{\rm max}=z'/\vpmo=(z-z_{d,\mathrm{max}})/\vpmo$.
Generally, $\vpmo$ can be values between $\varpi_d (1-\mu_0^2)$ and $\varpi_d$ 
(see discussions in Sec. \ref{sec:disk}), but in the fiducial model we present here,
$\vpmo$=$\varpi_d$, in which case, Eq. (\ref{eq:zdmax}) and (\ref{eq:Rmax}) become
\beq
z_{d,\mathrm{max}}=\varpi_d \mu_0 \left[ \left( \frac{1}{1-\mu_0^2} + \mu_0^2 \right)^{1/2} -1 \right]
\eeq
and
\beq
\rmax=\left[Z_{\rm max}^2\left(\frac{1-\mu_0^2}{\mu_0^2}\right)+
2Z_{\rm max}\left(\frac{1-\mu_0^2}{\mu_0}\right)+
2Z_{\rm max}\left(\frac{z_{d,\mathrm{max}}}{\varpi_d}\right)\left(\frac{1-\mu_0^2}{\mu_0^2}\right)+
1 \right]^{1/2}.
\eeq
For $\twe\simeq 60 ^\circ$, we find that $\eta$ is typically 
between 1 and 3, but can 
reach values as high as 7 close to the outermost streamline,
for the case considered in the text in which $\xmo$ is very large. 
Fully expanded
disk winds 
are asymptotically similar to X-winds (\citealt[]{Shu94}) and
have $\eta\ll 1$; such winds are possible only if $\xmo$ is not large.

\end{document}